\documentclass[fleqn,usenatbib]{mnras}

\usepackage{newtxtext,newtxmath}

\usepackage[T1]{fontenc}
\usepackage{ae,aecompl}

\usepackage{graphicx}	
\usepackage{amsmath}	

\usepackage{subcaption}
\title[FOBOS: Few Observation Binary Orbit Solver]{Few Observation Binary Orbit Solver ({\tt FOBOS}) from two (or more) astrometric observations}

\author[R. J. Houghton., S. P. Goodwin.]{
Rebecca J. Houghton,$^{1}$\thanks{E-mail: rhoughton1@sheffield.ac.uk (RJH)}, Simon P. Goodwin,$^{1}$
\\
$^{1}$Department of Physics and Astronomy, The University of Sheffield, Hounsfield Rd, Sheffield, S3 7RH, UK\\
}

\pubyear{2021}

\begin{document}
\label{firstpage}
\pagerange{\pageref{firstpage}--\pageref{lastpage}}
\maketitle

\begin{abstract}
We have developed a new, fast method of estimating the orbital properties of a binary or triple system using as few as two epochs of astrometric data. {\tt FOBOS} (Few Observation Binary Orbit Solver) uses a flat prior brute force Monte Carlo method to produce probability density functions of the likely orbital parameters. We test the code on fake observations and show that it can (fairly often) constrain the semi-major axis to within a factor of 2-3, and the inclination to within $\sim$20$^{\circ}$ from only two astrometric observations. We also show that the 68 and 95 per cent confidence intervals are statistically reliable. Applying this method to triple systems allows the relative inclination of the secondary and tertiary star orbits to be constrained. {\tt FOBOS} can usually find a statistically significant number of possible matches in CPU minutes for binary systems, and CPU hours for triple systems.
\end{abstract}



\section{Introduction}
\label{sec:introduction}

Many (probably the vast majority) of stars form in multiple systems \citep{DucheneKraus2013,ReipurthPPVI}, and so the properties of multiple systems (such as the semi-major axis distribution and relative inclinations of triple systems) contain a wealth of information on star formation \citep{Goodwin2010}.  Similarly, most stars seem to form planetary systems, and exoplanet orbits will contain information on the formation and dynamical evolution of planetary systems \citep{Winn2015}. Therefore, it is important and useful to constrain the orbital properties of stars and planets.

Orbital parameters can be found from observations covering multiple epochs of velocity and/or astrometric data. Several orbital fitting tools have been developed recently, including \texttt{BATMAN} \citep{BATMAN} and \texttt{RadVel} \citep{RadVel} (which are set up to use only transit light curves and radial velocity measurements respectively), as well as \texttt{orbitize!} \citep{Orbitize} and \texttt{ExoSOFT} \citep{ExoSOFT}.  

Unfortunately, what we most often have for the vast majority of multiple systems is a single epoch of observations from which extracting the orbital parameters of individual systems is impossible.  Potentially usefully, we may sometimes have a second epoch from follow-up observations.

We have developed a new orbital parameter finder - {\tt FOBOS} (Few Observation Binary Orbit Solver). {\tt FOBOS} is designed to find confidence limits for orbital parameters with only two epochs of observation. We will show that it is sometimes possible to strongly constrain the orbital parameters of binary or triple systems with only two epochs. {\tt FOBOS} can be used with a second epoch from follow-up observations, and we also hope it will act as an incentive to obtain a second epoch on what are currently single-epoch observations. {\tt FOBOS} is also extremely quick - often finding (sometimes quite tight) confidence limits for binary systems in a handful of CPU minutes, or triples in CPU hours. With more than two epochs of data {\tt FOBOS} can often become significantly more constraining.

In this paper, we describe the method used by {\tt FOBOS} and use fake observations of multiple systems to illustrate how well it can estimate orbital parameters. 

\section{Methods}
\label{sec:methods} 

{\tt FOBOS} uses a (flat prior) brute force Monte Carlo approach written in {\tt fortran90} and OMP parallelised to estimate the orbital parameters of binary and triple systems using as few as two observations of a system. As we will show, just two epochs of observations can sometimes tightly constrain orbital parameters in binary and triple systems, and three or more epochs can narrow these constraints even further. The majority of the testing in this paper is done on stellar systems, although in section \ref{sec:comparisons} we show that this approach also works for lower mass (brown dwarf) companions. For planetary triple systems, one would need to alter the stability condition used later for triple systems. 

The method works by generating fake systems with a random set of orbital parameters, projecting them into 2D, and comparing the positions of the companion star(s) at the different epochs to establish whether the orbital parameters of the fake system match the observations (to within the observational errors). We show that the error estimates are statistically reliable (ie. the actual parameters are within the 68 and 95 per cent confidence limits as often as would be expected). A full breakdown of how the code works is given in section \ref{subsec:setup}. 

An orbit is characterised by three unchanging physical parameters: the semi-major axis $a$, eccentricity $e$, and the inclination of system relative to the observer, $i$. There are also two `instantaneous' orbital properties: the phase of the orbit (the true anomaly), $\nu$, and the orientation of the system, $\phi$. 

\begin{figure}
    \centering
    \includegraphics[width=\columnwidth]{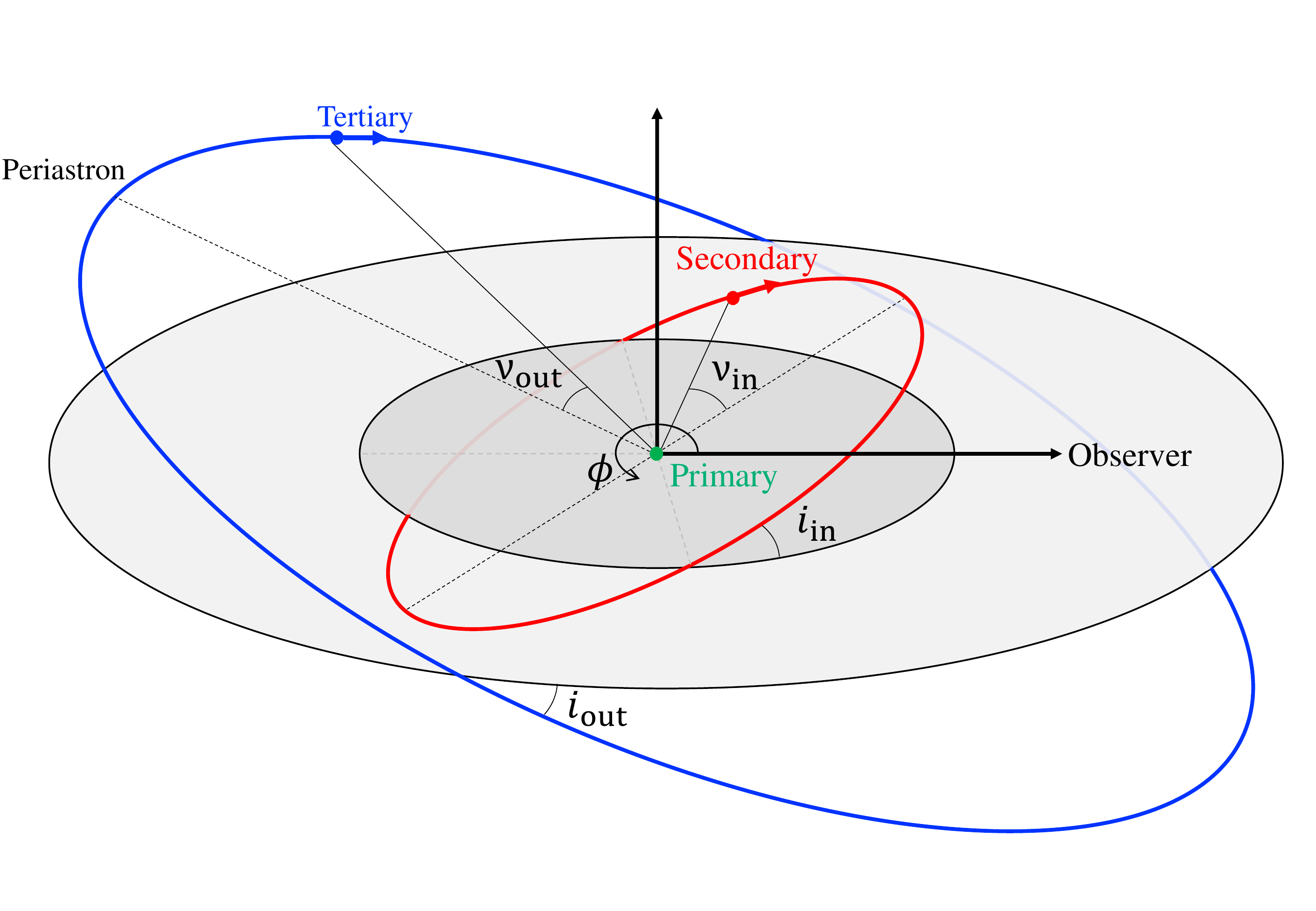}
    \caption{Diagram of a triple system. The diagram depicts the orbit of secondary and tertiary stars around a primary star. Each orbit is inclined with an inclination $i$ relative to an observer and rotated through an angle $\phi$ about an axis perpendicular to the line of sight. The true anomaly of each star is measured from periastron, assuming periastron is coincident with the line of sight for $\phi=0\degr$.}
    \label{fig:diagram}
\end{figure}

The true anomaly, $\nu$, we define such that when $\nu=0\degr$, the companion star is at periastron (i.e. at closest approach to the primary). Similarly, $\phi=0\degr$ is defined such that the semi-major axis of the system is along the line of sight. The definitions $i$, $\phi$, and $\nu$ are illustrated in Fig. \ref{fig:diagram}.

Often, stellar orbits will be parameterised by the longitude of periastron, $\omega$, and the longitude of the ascending node/node position angle, $\Omega$. As we are producing fake observations of physical systems, and are aiming to constrain only the physical parameters of the system, we use $\phi$ as a single orientation term instead of $\omega$ and $\Omega$. 

For any two epochs of observations at times $t_1$ and $t_2$, the secondary (or tertiary) stars are separated by a distance $s_1$ and $s_2$ (in au) from the primary, with (arbitrary) position angles $\theta_1$ and $\theta_2$, differing by an angle $\Delta \theta$. Note that $s_1$, $s_2$, and $\Delta \theta$ will have some observational uncertainty associated with them, and $s_1$ and $s_2$ in au depend on the distance and the uncertainty associated with it.

We assume that we can ignore the precession of an orbit, so the only parameter which is a function of time is the true anomaly, $\nu(t)$.  
The true 3D separation at a time $t$, $r(t)$, for a system with parameters $a$, $e$, and $\nu(t)$ is given by

\begin{equation}
    r(t) = \frac{a(1-e^2)}{1 + e\cos \nu(t)}.
    \label{eqn:rad}
\end{equation}

The on-sky separation at a given time, $s(t)$, is related to the true 3D separation, $r(t)$, and instantaneous angles $\nu(t)$, $\phi$, and $i$ via

\begin{equation} \label{eq:sep}
    s(t) = r(t) \; (1-\sin^2(\phi+\nu(t))\sin^2 i)^{1/2}.
\end{equation}

We also need the orbital period, $P$: 

\begin{equation}
    P=\sqrt\frac{a^3}{m_1+m_2}.
    \label{eqn:period}
\end{equation}

Note that to calculate the orbital period of the system, we also need to know the masses of the primary and companion stars (which may have significant observational uncertainties, particularly for lower mass companions).

Therefore, in the simplest case of two epochs of observations, we have a known time difference $\Delta t = t_2 - t_1$, a change of on-sky angle $\Delta \theta$, and two separations, $s_1$ and $s_2$. These are related to the change in $\nu$ and $s$ between observations which depend on $P$, $a$, $e$, $\phi$, and $i$.

\subsection{Parameter constraints}
\label{subsec:setup}

{\tt FOBOS} samples from uniform distributions of parameters without any other prior assumptions on the form of the semi-major axis or eccentricity distributions.

An absolute lower limit on the semi-major axis $a_{\mathrm{min}}$ is given by half of the projected on sky separation of the primary and companion star. This is because the true distance of the star has a maximum in a highly eccentric ($e \sim 1$) system, such that $r_\mathrm{max} \sim 2a$. If the system is inclined, we observe a projected separation $s$ that is almost always smaller than the true distance. Therefore, $a_\mathrm{min} = s/2$. As this method requires at least two observations that will usually have different separations, $a_{\mathrm{min}}$ is calculated using the largest value.

Another highly constraining feature of this method involves calculating the on-sky velocity of the star, $v_\mathrm{obs}$, based on the distance it has moved in the time between observations. The on-sky velocity is a lower limit on the star's true orbital velocity.

The companion star reaches it's maximum orbital velocity, $v_{\rm max}$, at periastron, so for an orbit with parameters $a$ and $e$ 

\begin{equation} \label{eq:vmax}
    v_\mathrm{max}=\sqrt{\frac{G(m_1+m_2)}{a} \frac{(1+e)}{(1-e)}}.
\end{equation}

Due to the fact that $v_\mathrm{obs}$ is a lower limit on the speed of the companion star, it is only possible for it to have orbital parameters that satisfy 

\begin{equation} \label{eq:amax}
    a < \frac{G (m_1 + m_2)}{v_{\mathrm{obs}}^2} \frac{(1+e)}{(1-e)}
\end{equation}

By assuming that it is extremely unlikely for our observed system to have an eccentricity of $e\gtrsim0.98$, eqn. \ref{eq:amax} can be used to give a probable upper limit on the semi-major axis of the companion
\begin{equation} \label{eq:amax2}
    a_\mathrm{max} = 100 \times \frac{G (m_1 + m_2)}{v_{\mathrm{obs}}^2} \; \; \; \mathrm{for} \; e<0.98.
\end{equation}

For systems with very large on sky velocities, this can be highly constraining. This gives a useful upper limit on the semi-major axis, as by reducing the possible range of parameter space to be sampled. 

In the event that the simulation manages to find no solutions, the limit on the semi-major axis can be removed to allow sampling of extremely high eccentricities at larger-$a$ than previously allowed.

Note that eqn. \ref{eq:amax} also contains the masses of the stars.  We use the upper limit on the masses to determine $a_\mathrm{max}$, as these give the largest possible value of $a_\mathrm{max}$.

\subsection{Orbital Parameter Generation}
\label{subsec:orbitgen}

At the beginning of each iteration of the Monte Carlo simulation, we select random values for each of the orbital parameters as described above, which are within the ranges shown in table \ref{tab:paramranges}.

\begin{table} 
    \centering
    \begin{tabular}{c c c c}
        \hline
        Parameter & Symbol & Range & Units \\
        \hline
        Semi-major axis & $a$ & $a_\mathrm{min}$ - $a_\mathrm{max}$ & au \\
        Eccentricity & $e$ & 0 - 1 & - \\
        Inclination & $i$ & 0 - 90 & deg \\
        Orientation & $\phi$ & 0 - 360 & deg \\
        Mean anomaly & $M$ & 0 - 360 & deg \\
        \hline
    \end{tabular}
    \caption{Definitions of the physical and instantaneous orbital parameters along with their allowed ranges.}
    \label{tab:paramranges}
\end{table}

In order to avoid any biases in the posterior PDFs, this method assumes flat uniform priors when selecting the semi-major axis, eccentricity, orientation, and mean anomaly values. The inclination is selected such that it is uniform in $\sin i$, meaning that $i$ is preferentially closer to edge-on (0$^\circ$) than face-on (90$^\circ$) (as would be expected from observing a random distribution of inclinations in 3D).

The true anomaly is generated from a distribution that is uniform in time. We calculate $\nu$ by first selecting a random value between 0 and $2\pi$ for the mean anomaly $M$ of the star. This is then converted to the true anomaly, $\nu$, by solving Kepler's equation using the Newton-Raphson method; this requires eqn. \ref{eq:eccan} to be solved numerically to find the value of the eccentric anomaly, $E$, via

\begin{equation} \label{eq:eccan}
    E - e\sin(E) = M,
\end{equation}

and then

\begin{equation} \label{eq:nu}
    \tan\left(\frac{\nu}{2}\right) =\sqrt{\frac{(1+e)}{(1-e)}}\tan\left(\frac{E}{2}\right).
\end{equation}

Once all 5 parameters have been selected/calculated for our test system, we can move on to producing a `fake' observation. 

We assume the set of parameters corresponds to the first observation at time $t_1$ and we rotate and project the system to find the separation $s(t_1)$.

If the test separation $s(t_1)$ does not match the observed first separation to within the observational errors the parameters are rejected as a possible match and we restart the process\footnote{It might seem that also testing if it fits $s(t_2)$ would be sensible, but this makes essentially no difference to the speed of the code as it makes the algorithm slightly more time-consuming.}.

If the test separation is a possible match to the observed system we can then proceed to advance the system forward in time. This is done by calculating the period $P$ of the orbit, then dividing the time between the epochs of observation by $P$ to calculate the fraction of an orbit through which the secondary star will move in time $\Delta t$. Since $M$ is uniformly distributed in time we can calculate $M_2$ at time $t_2$ from 

\begin{equation} \label{eq:meanf}
    M_2 = M_1 \pm \frac{2\pi\Delta t}{P}.
\end{equation}

Note that the companion could be moving in either direction around it's orbit, hence the $\pm$, and in elliptical orbits an equal change in $\pm M$ will almost certainly not correspond to an equal change in $\pm \nu$.  Note that the companion is allowed to have multiple orbits in time $\Delta t$ (which will occur if $\Delta t > P$). 

The two new values of the mean anomaly are converted to true anomalies using the same process as outlined above. These two new sets of parameters are projected onto the sky to see if either of the sets of $s(t_1)$, $s(t_2)$, and $\Delta \theta$ match their observed counterparts within the observational errors.

The final probability density function is calculated from all matches found for a particular set of observations (ideally at least 1\;000 matches, and never less than 300 - this is discussed further in section \ref{subsec:tripresults}). 

\subsection{A note on degeneracies}

As our observations are a projection onto the sky, the orientation and inclination are `degenerate'. The inclination may be such that the secondary is either in front of or behind the primary and we would have no way of knowing which. Therefore, an inclination of $20^{\circ}$ could correspond to either plus $20^{\circ}$, or minus $20^{\circ}$. Similarly, the orientation could be such that e.g. periastron was on the near side of the primary, or on the far side, and we would not be able to distinguish this. These degeneracies mean that it is often impossible to tell the direction of motion (e.g. clockwise vs. anticlockwise) of the orbit from only a two epochs (the exception would be an almost face-on orbit).  

For binaries, the fact that orientation, inclination, and direction are degenerate does not matter at all. However, in triple systems the degeneracy in inclination and the direction of the orbit can be important and will be discussed in section \ref{subsec:tripresults}.  

\subsection{Errors on observed quantities}
\label{subsec:compwobs}

The code compares the separations and position angles of the fake system to an observation. When running the code, a match will be triggered if both separations and the angle match within the observational errors. For the example systems tested in this paper, we apply a blanket error of 5 per cent to each separation and angle. This value was chosen as it represents an upper limit of typical observational errors. Unsurprisingly, smaller errors in the observation tend to tighten the constraints on a system while increasing the time to find solutions.

We assume that the possible true values of the observations fall uniformly within the the assigned observational errors. We could fold the observational errors more cleverly into the PDFs by weighting `hits' by their closeness to the observed values - however, while the confidence ranges we find for some systems can be really quite small, they are too large to justify the extra complexity of doing this.

We have assumed in our tests that observed systems will have a good {\em Gaia} distance available, or be within a cluster/star forming region with a good distance estimate. The distance can be included as an extra parameter to find the best fit for this as well.  If this is worth doing very much depends on how large the uncertainty in the distance is compared to the uncertainty in the angular separations and angular shift.

In our tests we also assume that the masses are known to a much greater accuracy than the uncertainty in the angular separations and angular shift, and so any error can be neglected.  This will often not be the case and the masses of the primary and companion(s) can be included as extra parameters to be sampled.  This will add computational expense as we now have two or three new parameters to include.

The impact of real observational errors (including the astrometric errors and errors on masses/distances) is discussed further in section \ref{sec:comparisons}.

\subsection{Selection effects}

In order to estimate orbital parameters {\tt FOBOS} requires an on-sky motion to be observed. 
Rather obviously, this means that if a system's orbital parameters are such that the companion's motion is too small to be observed we cannot estimate it's orbital parameters (other than extremely weak constraints based on it not being observed to move).

This means we are only able to estimate the orbital parameters of a biased subset of systems with the `right' orbital parameters. On a system-by-system basis this is not important - if a companion is observed to move we can obtain confidence limits on its orbital parameters. However, over a population of binary or triple systems we will miss particular configurations of parameters. We will address this in a future paper (in prep.) in which we examine populations and biases.

\section{Testing on binary systems}
\label{sec:testing} 

We tested {\tt FOBOS} on 60 fake observations of binary stellar systems. We show that we find the correct values for parameters within the 68 per cent and 95 per cent confidence intervals as often as we would expect. We also show that sometimes {\tt FOBOS} is surprisingly good at constraining orbital parameters (and when it cannot, it is statistically reliable in telling us so).

The orbital parameters, masses, and time between epochs for each of the synthetic binaries used to test our code are available online. The semi-major axis values are randomly distributed in the range 4-450 au and the other orbital parameters within the ranges shown in table \ref{tab:paramranges} for each system. The time between epochs for each of the systems is $\sim$2-12 yrs, and the masses of the primaries are $m_1=0.2-1.4\;M_\odot$ and of secondaries $m_2=0.016-0.7\;M_\odot$. 

The only constraint we apply on selecting binary systems to test is that the companion star must have moved a distance greater than 1 per cent of the initial separation $s_1$ between observations such that it's motion on the sky is clearly visible.  While it is possible to constrain orbital parameters from an observation of no apparent motion, these constraints are {\em extremely} weak (the main constraint is that the on-sky velocity is too small to have been observed which rules-out some, usually close, orbital configurations).

Each of the test systems ran on a 6 core / 12 thread CPU and the simulation ended when the number of possible matches exceeded 50\;000. The performance of the code is discussed in section \ref{subsec:timing}, but often solution PDFs can be found in minutes. 

We found that 45/60 (75 per cent) simulations correctly identified the semi-major axis of the binary within the 68 per cent confidence range, and 58/60 (97 per cent) within the 95 per cent confidence range. Similarly, the true inclination of the system is within the 68 per cent range for 41/60 (68 per cent) of test systems and 95 per cent confidence range for 57/60 (95 per cent) of systems. The eccentricity has 35/60 (58 per cent) and 59/60 (98 per cent) within the 68 per cent and 95 per cent confidence intervals respectively.  

The key point here is that {\tt FOBOS} gets the `wrong' answer as often as one would expect.

\subsection{General performance}

We find that {\tt FOBOS} is often good at constraining orbital parameters, with the eccentricity being the most difficult parameter to constrain. Typically, we find that {\tt FOBOS} is able to indicate if the eccentricity is likely to be 'low', 'intermediate', or 'high'. This can be seen from the from the full table of 68 and 95 per cent confidence intervals (for all 60 test systems) that is available online. 

The 68 per cent confidence limits on the semi-major axis are often within a factor of $<3$ (21/60 systems), mostly within a factor of 5 (40/60 systems), and in only 2 cases a factor of 10 or more. Given the difficulty in constraining eccentricity there is usually a `floor' of a factor of 2 on constraining the semi-major axis.

{\tt FOBOS} is often very good at constraining the inclination of the system - in 26/60 systems the 68 per cent confidence limits are less than $20^{\circ}$, and only 1/60 is beyond $40^{\circ}$.

\subsubsection{System B17}

\begin{figure}
    \centering
    \includegraphics[width=\columnwidth]{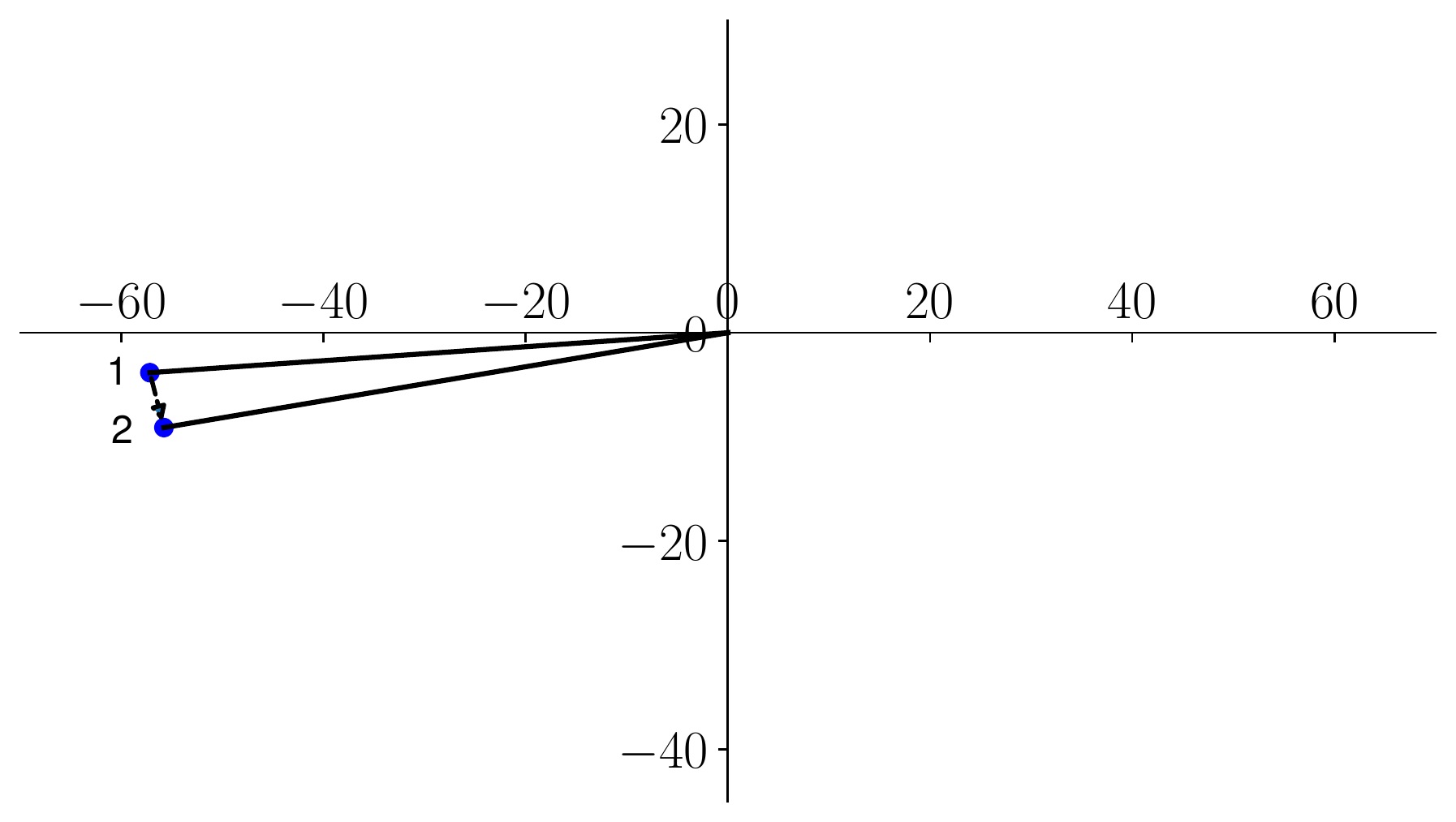}
    \caption{On sky projection of system B17 at two epochs. The position of the companion star at the first and second epochs of observation are marked as 1 and 2 respectively, with the direction of the star's on sky motion shown by the arrow. The primary star is located at (0,0) in both observations. The axes are in au.}
    \label{fig:projection3}
\end{figure}

\begin{figure*}
    \centering
    \includegraphics[width=.95\textwidth]{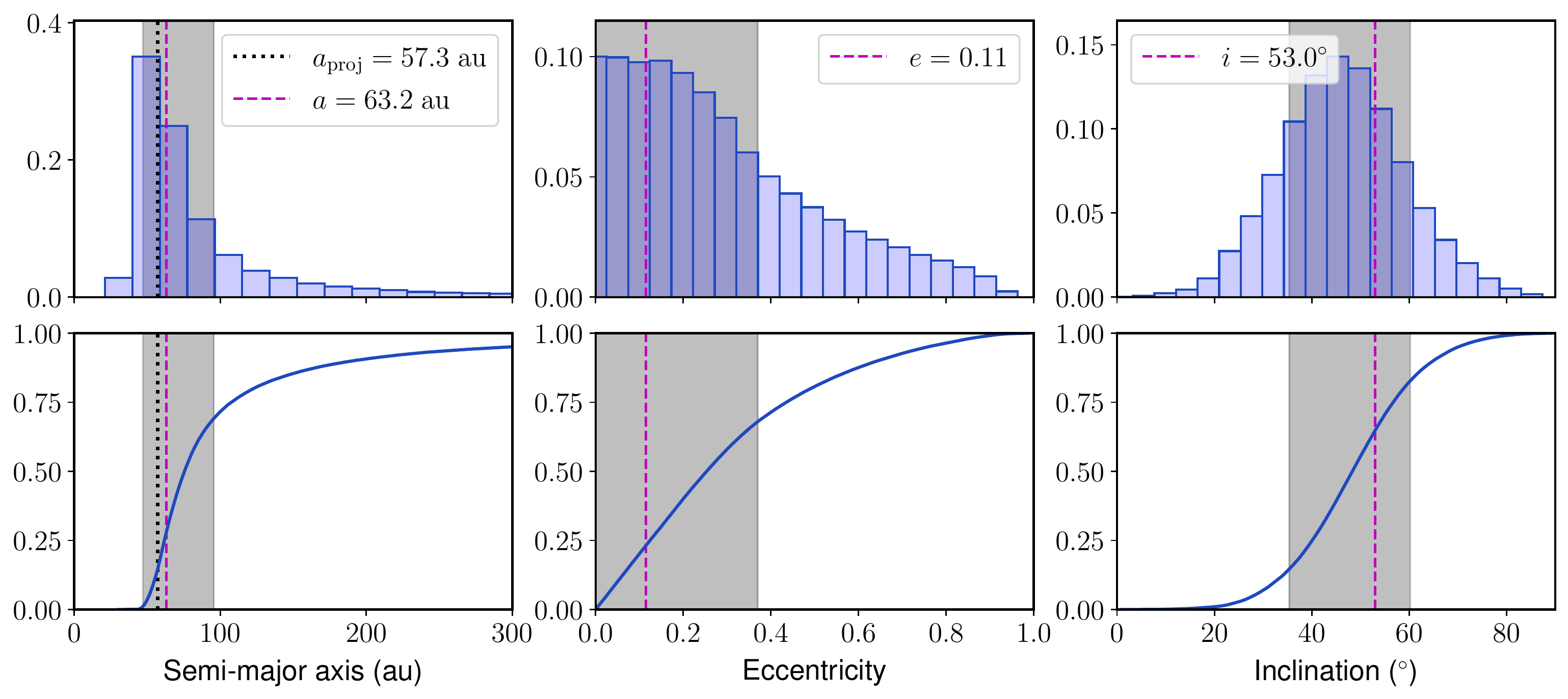}
    \caption{The final probability density functions for system B17 show as a histogram (top) and a cumulative distribution function (bottom) for the semi-major axis (left), eccentricity (middle), and inclination (right). The true parameters for the semi-major axis, eccentricity, and inclination are shown as the magenta dashed lines on each plot, and the minimum projected separation of the two stars (in au) is shown by the black dotted lines on the semi-major axis plots. The shaded regions represent the 68 per cent confidence intervals on each of the orbital parameters.}
    \label{fig:bin3-best}
\end{figure*}

An example of the ability of {\tt FOBOS} to find tight constraints on orbital parameters is system B17. Fig. \ref{fig:projection3} shows the on sky projection of system B17 at the two epochs.  Note that the position angles are completely arbitrary - only the change in position angle, $\Delta \theta$, is important.

This binary system has a maximum projected separation of $s=57.3\;\mathrm{au}$, meaning that the lower limit on the semi-major axis is $a_\mathrm{min}=28.6\;\mathrm{au}$. The time between observations was $7.43$ years, during which the star moved a distance of $5.45\;\mathrm{au}$ on the sky. Therefore, the observed on sky velocity of the star was $0.73\;\mathrm{au\;yr^{-1}}$, or $v_\mathrm{obs}=3.49\;\mathrm{km\,s^{-1}}$. The velocity gives an upper limit of $a_\mathrm{max}=7\,930\;\mathrm{au}$ to the semi-major axis using eqn. \ref{eq:amax2}. This upper limit is for the extreme case of the system being observed face-on while the companion is at periastron in a very highly eccentric orbit.  (Note that we will usually quote results to three significant figures, for real data this should obviously depend on the relative size of the errors on various quantities.)

Fig. \ref{fig:bin3-best} shows the resulting probability density functions for semi-major axis (left), eccentricity (middle), and inclination (right) - as a histogram (top), and CDF (bottom). The 68 per cent confidence ranges are shown by the grey shaded regions and the true value of the semi-major axis, eccentricity, and inclination are shown by the purple dashed-lines in each panel.  For the semi-major axis the black dotted line shows the maximum observed separation.

In this case, {\tt FOBOS} has performed extremely well. The 68 per cent confidence limits for $a$ are $47.0-95.6$ au (true value 63.2 au), for $e$, $0.00-0.37$ (true value 0.12), and for $i$, $35.4-60.2^{\circ}$ (true value $53.0^{\circ}$).

\begin{figure*}
    \centering
    \includegraphics[width=\textwidth]{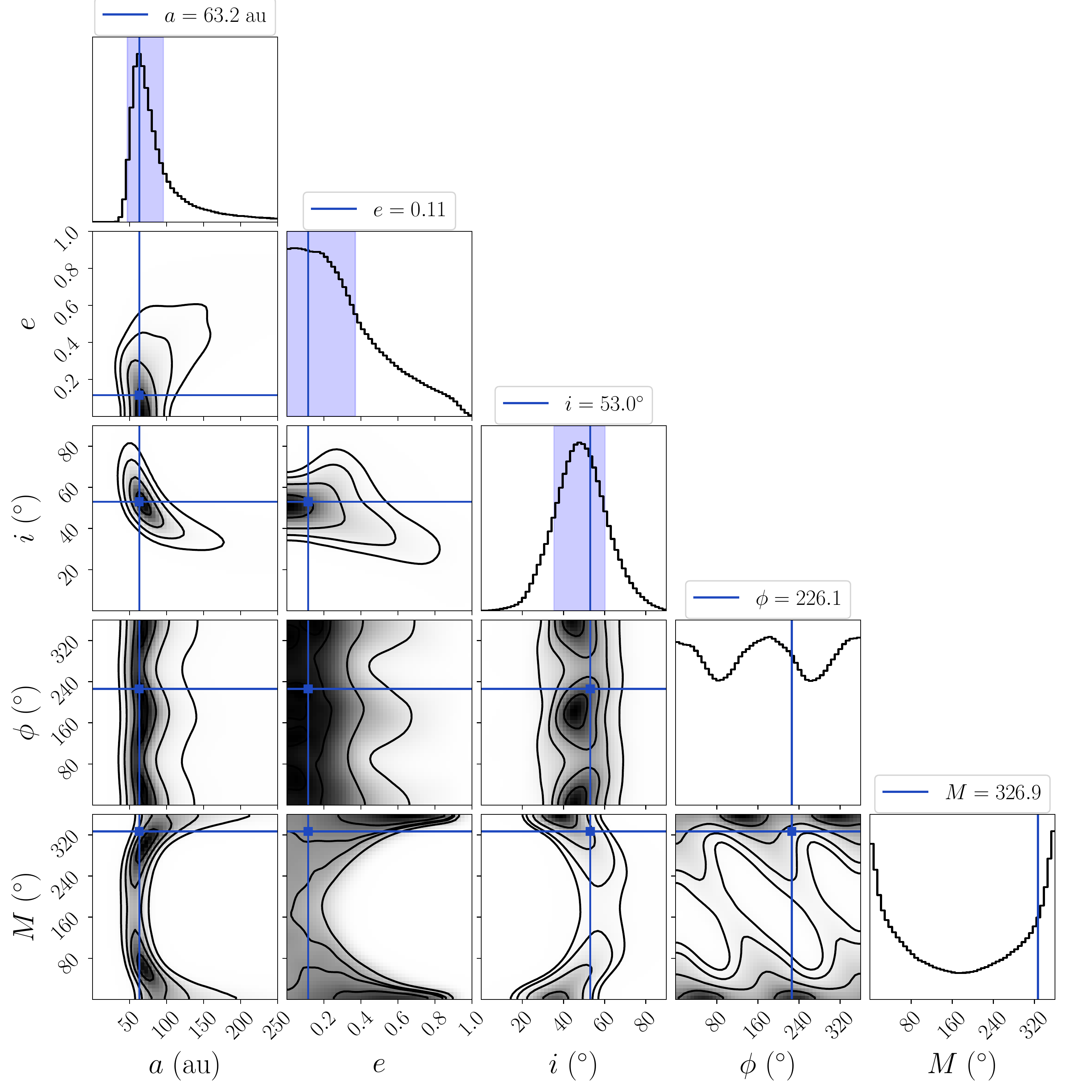}
    \caption{Corner plot of parameters $a$, $e$, $i$, $\phi$, and $M$ for system B17, with solid blue lines representing the true values and the shaded blue regions showing the {\tt FOBOS} 68 per cent confidence intervals. The panels at the top of each column show the probability density functions for each orbital parameter individually and all other panels show two dimensional covariance of each combination of parameters.}
    \label{fig:bin-corner3}
\end{figure*}

Corner plots are useful to examine the connection between different parameters.  In Fig. \ref{fig:bin-corner3} we show the corner plot for system B17 - note that as well as $a$, $e$, and $i$, {\tt FOBOS} can also estimate the instantaneous orbital parameters $\phi$ (orientation), and $\nu$ or $M$ (phase).  

Fig. \ref{fig:bin-corner3} shows slightly more subtle information than the individual PDFs in Fig. \ref{fig:bin3-best}.  Semi-major axis and eccentricity are (unsurprisingly) related, and we can see that if $a$ is high, then $e$ must be high (far left, second panel down). The orientation ($\phi$, forth row) of the orbit shows a slight preference for being close to either $\phi=0^\circ$ or $\phi=180^\circ$, but could take any value in the $0-360^\circ$ range. The phase ($M$, bottom row), however, is well constrained to be probably very close to periastron ($M \sim 0^{\circ}$). Depending on what one is interested in in a particular system the instantaneous orbital parameters may be extremely interesting or of little use.

The information in the corner plot can allows us to rule-out particular combinations of parameters in a way that is not obvious from the individual PDFs.  For example, if we were to have extra information that made us suspect that $a$ was high (say, $>200$ au) then that would constrain $e$ to being high ($>0.4$), and $i$ to be quite low ($< 50^{\circ}$).

\subsubsection{System B4}

\begin{figure}
    \centering
    \includegraphics[width=\columnwidth]{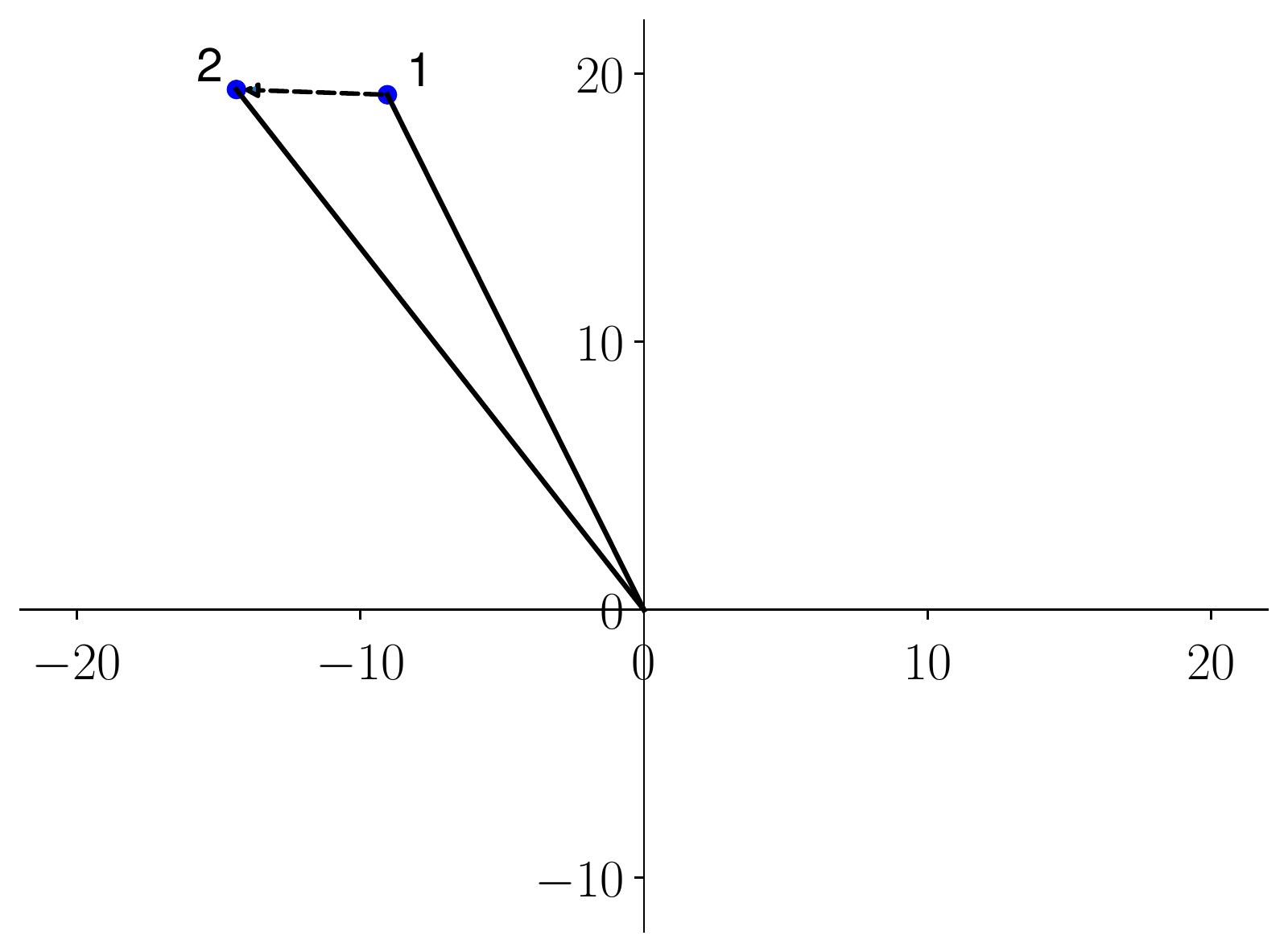}
    \caption{On sky projection of system B4 at two epochs. Annotations and axis units are as in Fig. \ref{fig:projection3}.}
    \label{fig:projection4}
\end{figure}

A much less well constrained system is system B4 whose observation is shown in Fig. \ref{fig:projection4}. It is worth comparing the observations of systems B17 and B4 in Figs. \ref{fig:projection3} and \ref{fig:projection4}. System B4 has moved slightly further than system B17 and the two observations appear to the eye as if they are very similar and contain very similar information. However, as we will see, the data for system B4 is not particularly constraining.

System B4 has true values of $a=190$ au, $e=0.51$, and $i=10.4^{\circ}$.  The minimum semi-major axis was calculated as $a_\mathrm{min}=12.1\:\mathrm{au}$ from an on-sky separation of $s_\mathrm{proj}=24.2\:\mathrm{au}$. The distance moved by the star in $10.6$ yrs corresponds to an on sky velocity of $v_\mathrm{obs}=2.39 \; \mathrm{km\,s^{-1}}$. These values do not appear to be dissimilar to other test systems.

\begin{figure*} 
    \centering
    \includegraphics[width=0.95\textwidth]{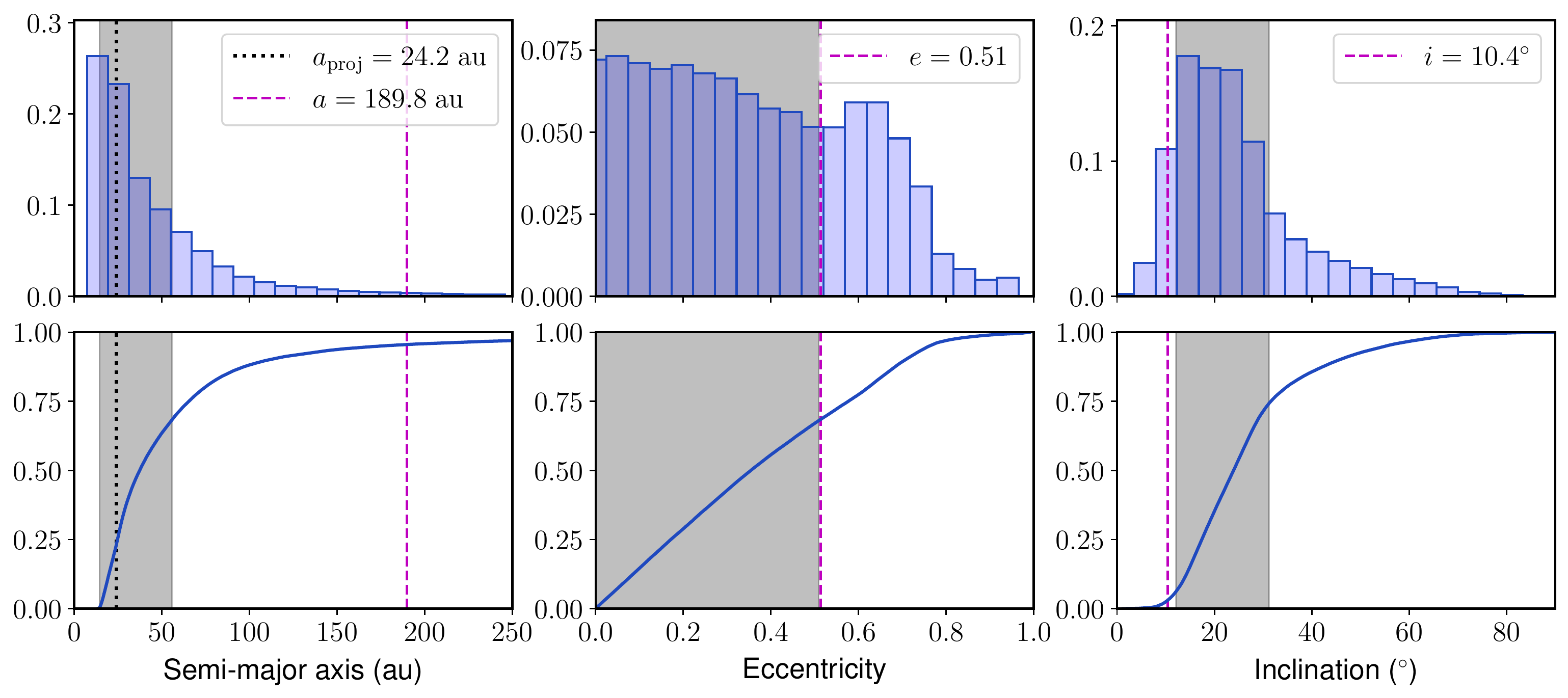}
    \caption{Probability density functions for system B4. Legend as in Fig. \ref{fig:bin3-best}.}
    \label{fig:bin4-worst}
\end{figure*}

The confidence limits for system B4 are shown in Fig. \ref{fig:bin4-worst}. Starting with the middle and bottom panels: the eccentricity and inclination are almost in the 68 per cent confidence limits. The inclination is fairly well-constrained as probably $10-30^{\circ}$. The eccentricity is probably less than 0.8, but the exact value would be difficult to estimate\footnote{The confidence limits are found by finding the smallest range of parameter values containing 68 and 95 per cent of the PDFs.  This fits peaks well, but in the case of the eccentricity distribution here, it doesn't quite map onto the almost flat PDF from 0 to 0.8.  This illustrates the usefulness of `eyeballing' PDFs.}.

However, we fail to correctly find the true semi-major axis of 190 au with a 68 per cent confidence range of $14-56$ au. The true value is just outside the 95 per cent confidence range of $14-174$ au. The corner plot for system B4 is included in the appendix (Fig. \ref{fig:bin-corner4}).

It should be noted that the code has not `failed' - it is just that of all the orbital parameters that could have produced the observed movement on the sky within the assumed errors, there were many with much smaller semi-major axes than what we know to be the actual answer.  The results are purely probabilistic and need to be treated as such: there is a higher probability that this particular projection of the motion of the binary on the sky corresponds to a system with a low eccentricity and small semi-major axis, rather than a relatively eccentric $e\sim0.5$ system with instantaneous orbital parameters that cause the projected separation of the stars to be eight times lower than the semi-major axis. 

\section{Triples}
\label{sec:triples}

The method outlined above can also be applied to hierarchical triple systems.  Hierarchical triples are composed of an inner binary and a significantly more distant outer tertiary companion. Therefore, we can consider a system as being composed of two independent orbits - the secondary star around the primary (referred to as the inner orbit) and the tertiary around the primary (outer orbit).  In hierarchical triples there needs to be a significant separation between the inner and outer orbits for the system to be stable which we show below is a very useful constraint.

For triple systems, we first assume that the star closest to the primary on the sky is the secondary star and the star furthest from the primary on the sky is the tertiary star. This is true for the majority of observations, but in some cases the tertiary star may appear closer to the primary than the secondary\footnote{Only in close-to edge-on systems for a small fraction of its orbit does the tertiary have the chance to be closer in the sky to the primary than the secondary.  One interesting case where this may become moderately likely is a system with a close-to face-on secondary and a close-to edge-on tertiary near the stability limit.}.  In cases where no fits can be found assuming the most probable alignment, it is possible to relax this assumption.

Each orbit will have it's own set of parameters, defined in the same way as for a binary. We use $a_\mathrm{in}$, $e_\mathrm{in}$, $i_\mathrm{in}$, $\phi_\mathrm{in}$, and $M_\mathrm{in}$ to denote the parameters of the inner orbit and $a_\mathrm{out}$, $e_\mathrm{out}$, $i_\mathrm{out}$, $\phi_\mathrm{out}$, and $M_\mathrm{out}$ for the outer orbit. These orbital elements are shown on the diagram in Fig. \ref{fig:diagram}. 

For systems with two companions, the inclination can vary from $-90\degr$ to $+90\degr$ as one orbit may be inclined above the plane on the side of the observer, and the other below. 

Attempting to fit five additional orbital parameters means that simulations of triple systems are significantly more computationally expensive. However, we can significantly reduce parameter space by excluding all unstable systems. 

\subsection{Stability}
\label{subsec:stability}

The stability of a triple system is determined by the semi-major axes, eccentricities, and the relative inclinations of the secondary and tertiary. 
There is no single empirical stability equation for hierarchical triple systems, although there are several widely used models including \cite{Harrington1972}; \cite{EggletonKiseleva1995}; \cite{valtonen2008}; \cite{ReipurthMikkola2012}. One of the most commonly used stability equations is the criteria of \cite{MardlingAarseth1999}, shown in eqn. \ref{eq:stab-mardaars99}, derived based on the chaotic energy and angular momentum interactions between the orbits of the two stars

\begin{equation} \label{eq:stab-mardaars99}
    \begin{aligned}
        \frac{a_\mathrm{out}}{a_\mathrm{in}}|_{\mathrm{crit}} = \frac{2.8}{1-e_\mathrm{out}} \left(1-\frac{0.3i_{\rm rel}}{\pi}\right)
        \left( \frac{(1.0+q_\mathrm{out})(1+e_\mathrm{out})}{\sqrt{1-e_\mathrm{out}}} \right)^\frac{2}{5},
    \end{aligned}
\end{equation}
where $e_\mathrm{out}$ is the eccentricity of the outer star, and $i_{\rm rel}$, is the relative inclination between the inner and outer orbits, and 
\begin{equation} \label{eq:massratio}
    q_\mathrm{out}=\frac{m_3}{m_1+m_2},
\end{equation}
where $m_1$, $m_2$ and $m_3$ are the masses of the primary, secondary, and tertiary stars respectively. A system is unstable if 
\begin{equation} \label{eq:stabilityratio}
    \frac{a_\mathrm{out}}{a_\mathrm{in}} > \frac{a_\mathrm{out}} {a_\mathrm{in}} |_{\mathrm{crit}},
\end{equation}
i.e. the ratio of the outer semi-major axis to the inner semi-major axis must be greater than the critical value given by eqn. \ref{eq:stab-mardaars99}.

This stability condition is valid for stellar mass objects, and for prograde orbits. It also ignores a small dependence on the inner mass ratio and inner eccentricity. However, it provides a conservative estimate of the stability of an orbit, occasionally rejecting stable orbits in order to ensure no unstable orbits are accepted.

\subsubsection{Generating fake triples}

The code treats a triple system as two individual orbits. In both cases, the primary star is at the centre of our co-ordinate system. Each orbit is modelled through the same process that is described in detail in section \ref{sec:methods}, the first stage of which is generating and projecting the inner orbit for both epochs. 

If both separations and the difference in position angle match the observation of the secondary star, then the simulation moves on to the outer orbit. We calculate a lower limit on $a_\mathrm{out}$ by evaluating eqn. \ref{eq:stab-mardaars99} for the selected values of $a_\mathrm{in}$ and $e_\mathrm{out}$, this ensures that all fake systems would be (hypothetically) stable. 

The vast majority of iterations end without finding a match for the inner orbit (full details of the rejection rate for various test systems is explored in section \ref{subsec:timing}). When a match is found for the inner orbit, 1 000 orbital configurations for the outer orbit are sampled to look for possible matches. 

\subsection{Results}
\label{subsec:tripresults}

The code was tested on 60 fake triple systems. Each simulation ran until 1\;000 matches had been found or the wall-clock time of the simulation exceeded 24 hours. The cutoff of 24 hours per simulation was an arbitrary time limit to ensure all simulations ran in a reasonable time frame, and should not be used for real systems.

Out of these 60 simulations, 4 of them (T14, T35, T44, T47) found between 300 and 1\;000 matches, and a further 6 simulations (T5, T18, T42, T46, T50, T56) produced fewer than 300 matches. These last 6 systems are excluded from the following statistics, as there were too few solutions to generate reliable probability density functions.

In tests it was found that 300 is an absolute lower limit on the number of matches required to have a statistically reliable probability density function, and when analysing real systems we would ideally want 1\;000 (or more) matches.

The true parameters for all of our triple systems are available in the online supplementary data. Note that the secondary and tertiary inclinations are both selected relative to the plane of the sky - in triple systems a much more useful and interesting measure is the relative inclination of the two orbits.

The semi-major axis, eccentricity, and inclination of the inner orbit were all within the 68 per cent confidence interval for 44/54 (81 per cent), 35/54 (65 per cent) and 38/54 (70 per cent) of systems respectively.  For the outer orbit these values are 36/50 (67 per cent), 32/54 (59 per cent) and 46/54 (85 per cent) respectively. 

{\tt FOBOS} is usually more effective at constraining the orbital parameters in triples compared to binaries due to the stability condition ruling-out many possible configurations which could otherwise fit the observations.

\subsubsection{System T19}

\begin{figure*}
    \centering
    \begin{subfigure}{\textwidth}
        \centering
        \caption{Secondary}
        \includegraphics[width=.95\textwidth]{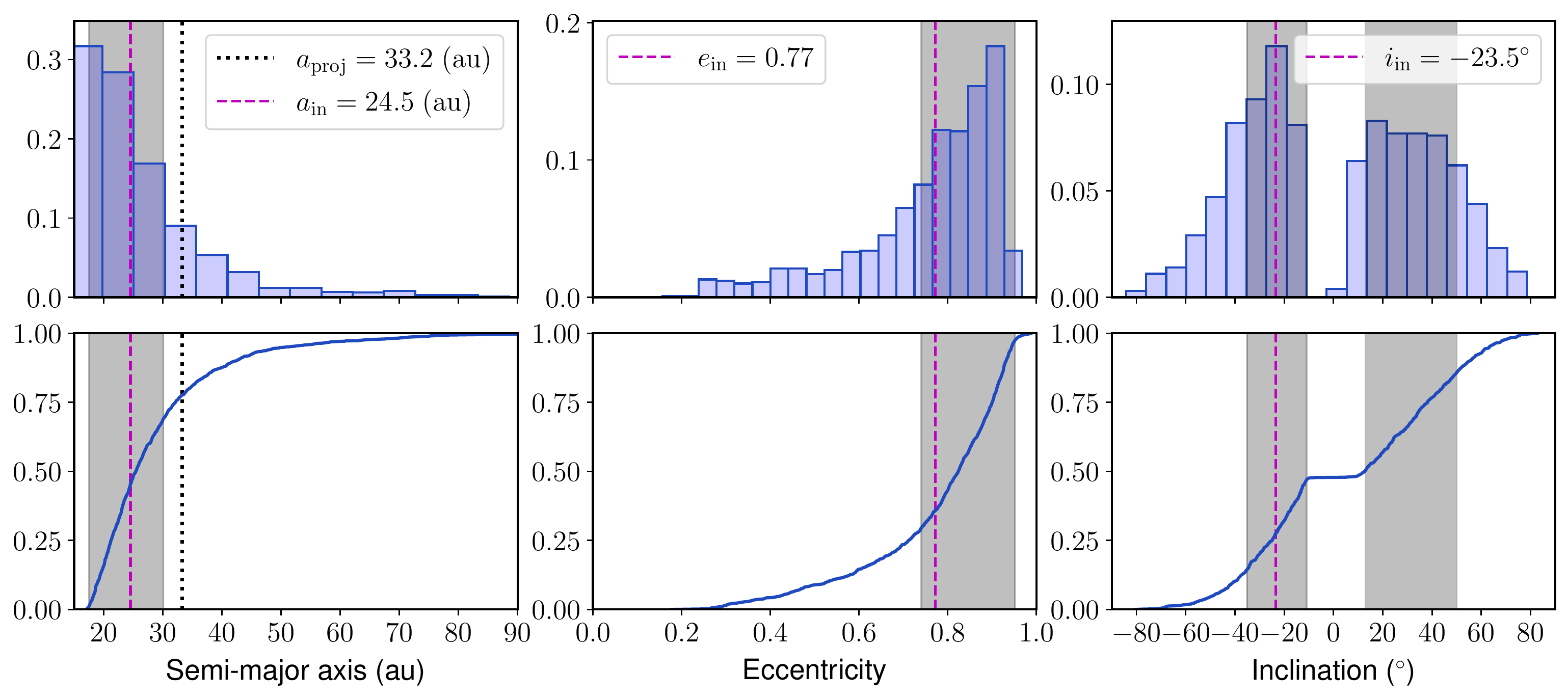}
        \label{fig:trip-19a}
    \end{subfigure}%
    \\
    \begin{subfigure}{\textwidth}
        \centering
        \caption{Tertiary}
        \includegraphics[width=.95\textwidth]{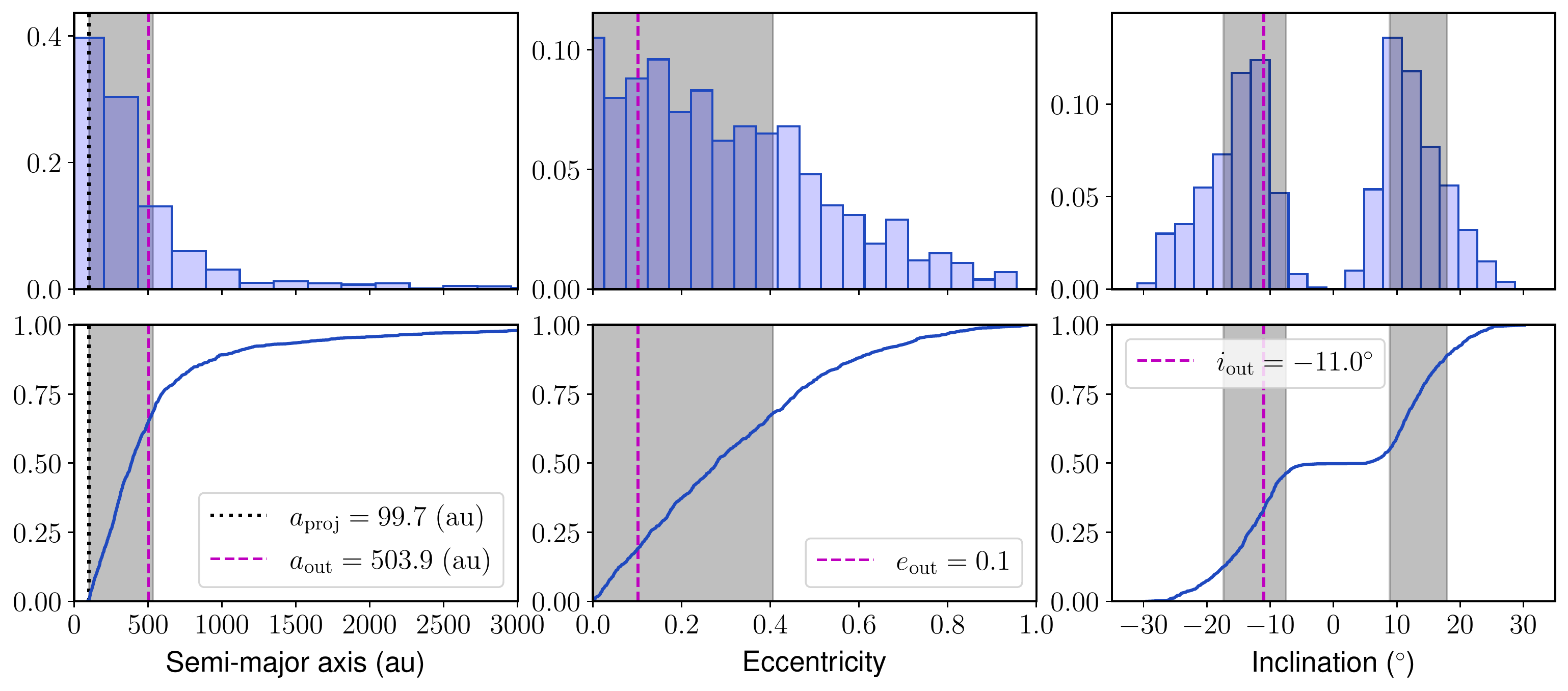}
        \label{fig:trip-19b}
    \end{subfigure}
    \caption{Histograms and cumulative distribution functions for semi-major axis, eccentricity, and inclination for the secondary (`a', top), and tertiary (`b', bottom) stars system T19. The grey shaded regions show the 68 per cent confidence interval for each parameter. The magenta dashed lines represent the true value of each orbital parameter and the black dotted line shows the maximum projected separation of the primary and secondary star out of the two observations.}
    \label{fig:trip-19}
\end{figure*}

System T19 is an example of a well constrained triple system.  The true parameters are $a_{\rm in} = 24.5$ au, $e_{\rm in} = 0.77$, and $i_{\rm in} = -23.5^{\circ}$, and $a_{\rm out} = 504$ au, $e_{\rm out} = 0.10$, and $i_{\rm out} = -11.0^{\circ}$. The relative inclination of the two orbits is $13.5^{\circ}$. The system was observed at two epochs which were 8.67 yrs apart.

The maximum projected separations of the secondary and tertiary stars were $31.2\:\mathrm{au}$ and $99.0\:\mathrm{au}$ respectively, and they  moved with on-sky velocities of $1.69\:\mathrm{km\,s^{-1}}$ and $1.73\:\mathrm{km\,s^{-1}}$. 

Fig. \ref{fig:trip-19} shows the PDFs of the secondary (top) and tertiary (bottom) for the semi-major axis (left), eccentricity (middle), and inclination (right).  Again, the shaded regions are the 68 per cent confidence ranges, the purple dashed lines give the true value, and the green dotted line in the top panels the maximum observed separation.  Note that the scales for semi-major axis and inclination are different for the secondary and tertiary.

The true semi-major axes of both the secondary and tertiary are within the 68 per cent confidence limits (left panels).  Interestingly, the semi-major axis of the secondary is found to be almost certainly significantly smaller than its projected separation; and the semi-major axis of the tertiary as almost certainly much larger than its projected separation.  Here the stability criterion is extremely powerful - if both the inner and outer semi-major axes of the components were close to their projected values the system would not be stable, hence the code has to move them in and out respectively to find mutually agreeable fits.

The eccentricities are fairly well constrained (middle panels).  The secondary eccentricity must usually be high to see the observed velocity shift for a low semi-major axis.  The tertiary eccentricity cannot be too high to fit the stability criteria (roughly speaking, the tertiary periastron needs to be at least about four times the secondary apastron), but is relatively weakly constrained as being probably less than 0.4. 

Note that the inclinations in the right panels are different to those used for binary orbits.  In binary orbits the inclination is given as a PDF between $0^{\circ}$ and $90^{\circ}$ as the degeneracy between e.g. $+45^{\circ}$ and $-45^{\circ}$ is unimportant.  However, in triple systems this degeneracy can be extremely important as it reflects the relative inclination of the companion stars.

The inclination distributions (the right panels of Fig. \ref{fig:trip-19}) both show two peaks which are roughly symmetric around zero degrees. This is because it is roughly equally likely to find solutions at plus or minus a particular inclination (the only difference being if the companion is in front of or behind the primary). The slight discrepancy between the confidence intervals at positive and negative inclinations is due to Poisson noise.  There is a relative inclination term in the stability condition (eqn. \ref{eq:stab-mardaars99}) which makes a slight difference to the symmetry, but this term is only important if a system is very close to the stability limit.

In the right panels of Fig. \ref{fig:trip-19} we can see that for the tertiary the inclination is well constrained at $\pm 8-18^{\circ}$, whilst the secondary is slightly less well constrained at $\pm 11-42^{\circ}$ (68 per cent confidence limits). The quoted confidence intervals are calculated assuming the inclinations are symmetric about  zero (which is usually the case).

It is worth mentioning that the relative directions (prograde or retrograde) of the orbit could provide extra information if they were available. If the inclination is constrained to be close-to face-on then the direction of the orbit can be determined. However, in the much more common case of close-to edge-on orbits relative directions cannot be determined\footnote{If both stars move in the same direction on the sky (e.g. left to right) they may have prograde orbits if they are both on the same side of the primary relative to us, or retrograde orbits if they are on opposite sides.  Unfortunately, from purely astrometric data we have no way of determining which side of the primary each companion is.  Additional radial velocity data could break this degeneracy, but we assume all we have is astrometric data.}.

There are two possible relative inclinations: one in which the relative inclination is small ($0-20^{\circ}$ if both are positive or both negative), or quite large ($20-60^{\circ}$ if they are opposite signs).  It is impossible to know which of these is true for an observed system (in system T19 we know that the correct answer is that the relative inclination is small).

We do potentially have a prior expectation in real systems that the formation mechanism (e.g. disc fragmentation) should produce triples which have similar inclinations.  With a population of real systems in which many have one possible configuration which is closely aligned in inclination we could make statistical/physical arguments for one configuration being more likely than the other.  However, in any single system considered in isolation it is impossible to distinguish.

\begin{figure*}
    \centering
    \includegraphics[width=\textwidth]{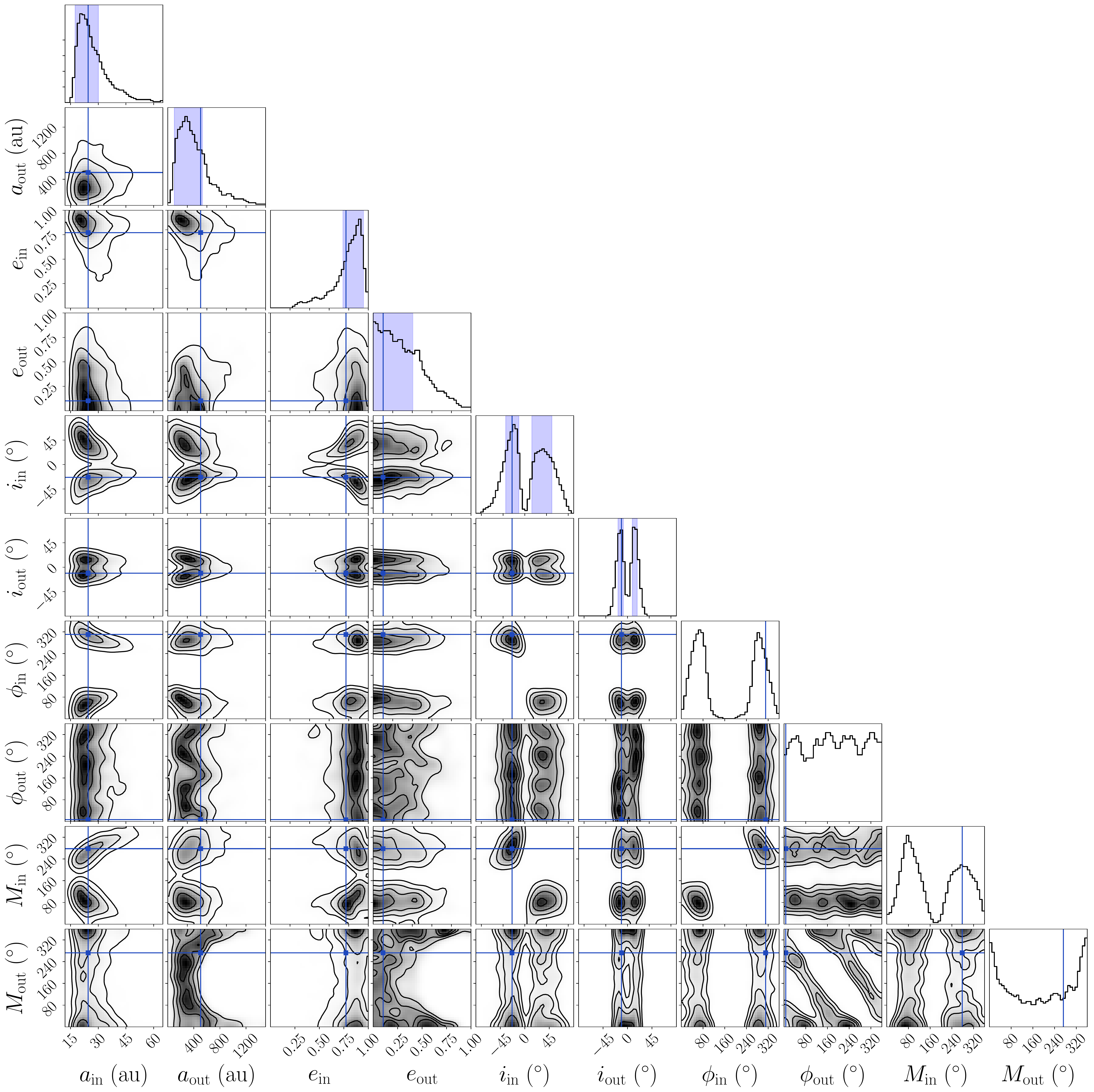}
    \caption{Corner plot for triple system 19, with solid blue lines representing the true orbital parameters (available in online data) and the blue shaded regions showing the {\tt FOBOS} 68 per cent confidence intervals}. Sample size of 1000 matches.
    \label{fig:corner-trip19}
\end{figure*}

We show the corner plot for system T19 in Fig. \ref{fig:corner-trip19}.  This is a much `busier' plot than for a binary system as we have many more parameters all of which are related to each-other.  Depending on what exactly one is interested in about a particular system, different parts of this plot will be more or less useful.  For example, the orientation, $\phi_\mathrm{in}$, of the inner orbit is very well constrained to be  around 70 or $290^{\circ}$ (these are symmetric, the difference being if periastron is in front or behind the primary).  This might be very useful information on the system (or not).

\subsubsection{System T25}

\begin{figure*}
    \centering
    \begin{subfigure}{\textwidth}
        \centering
        \caption{Secondary}
        \includegraphics[width=.95\textwidth]{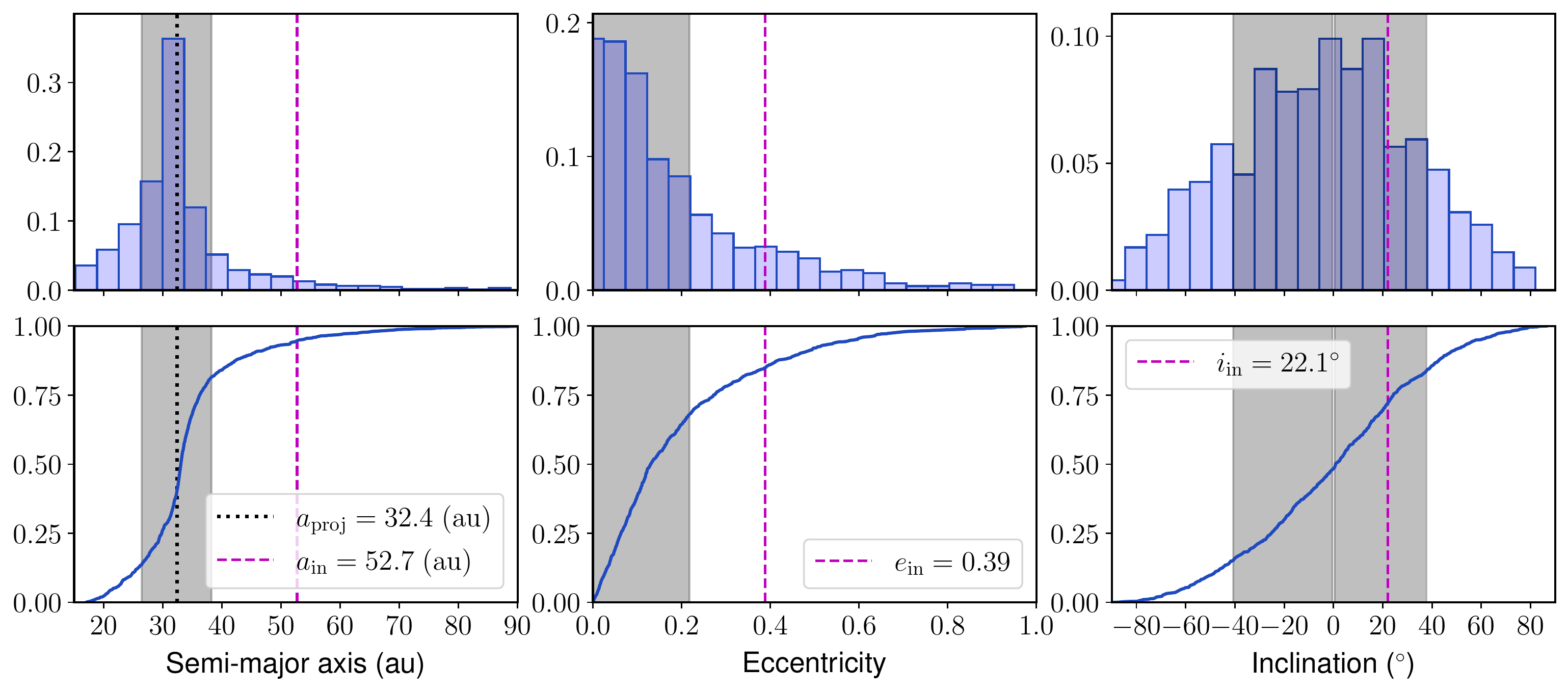}
        \label{fig:trip-25a}
    \end{subfigure}%
    \\
    \begin{subfigure}{\textwidth}
        \centering
        \caption{Tertiary}
        \includegraphics[width=.95\textwidth]{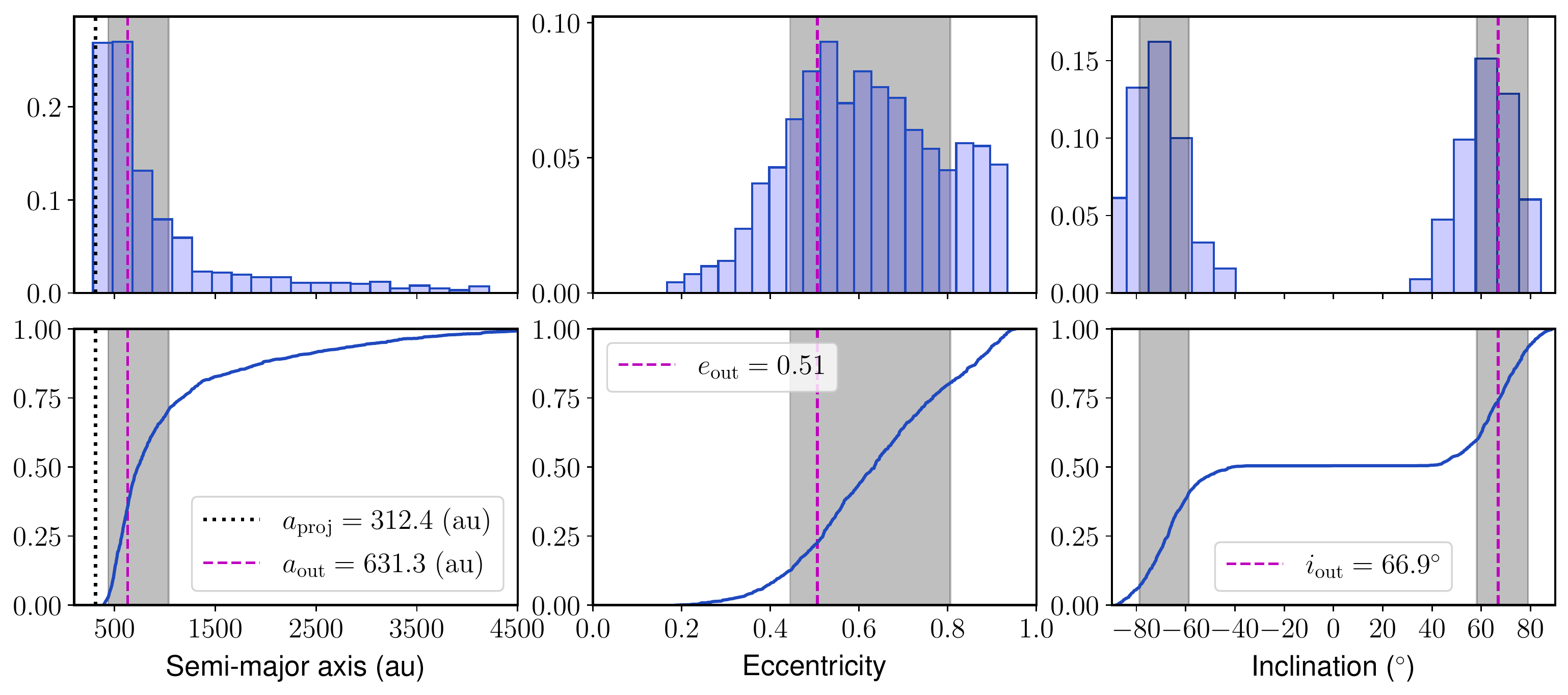}
        \label{fig:trip-25b}
    \end{subfigure}
    \caption{Histograms and cumulative distribution functions for semi-major axis, eccentricity, and inclination for the secondary ('a', top) and tertiary ('b', bottom) stars in  system T25. Legend as in Fig. \ref{fig:trip-19}.}
    \label{fig:trip-25l}
\end{figure*}

For system T25 we show the semi-major axis, eccentricity, and inclination PDFs for the secondary and tertiary in Fig. \ref{fig:trip-25l}.  System T25 shows some interesting features. The semi-major axis histogram shows a sharp peak centred on the projected separation of the secondary, whilst the true value lies outside the 68 per cent confidence interval and barely within the 95 per cent confidence interval. Also, the PDF for the inclination of the system does not show the same bimodality as the vast majority of the other systems, as we cannot constrain the values at all well, and the 68 per cent confidence interval is very large (essentially, the code cannot fit close-to face-on orbits, but anything less than about $\pm 45^{\circ}$ has a roughly equal probability).  However, it does a remarkably good job of constraining the tertiary orbit.

\begin{figure}
    \centering
    \includegraphics[width=\columnwidth]{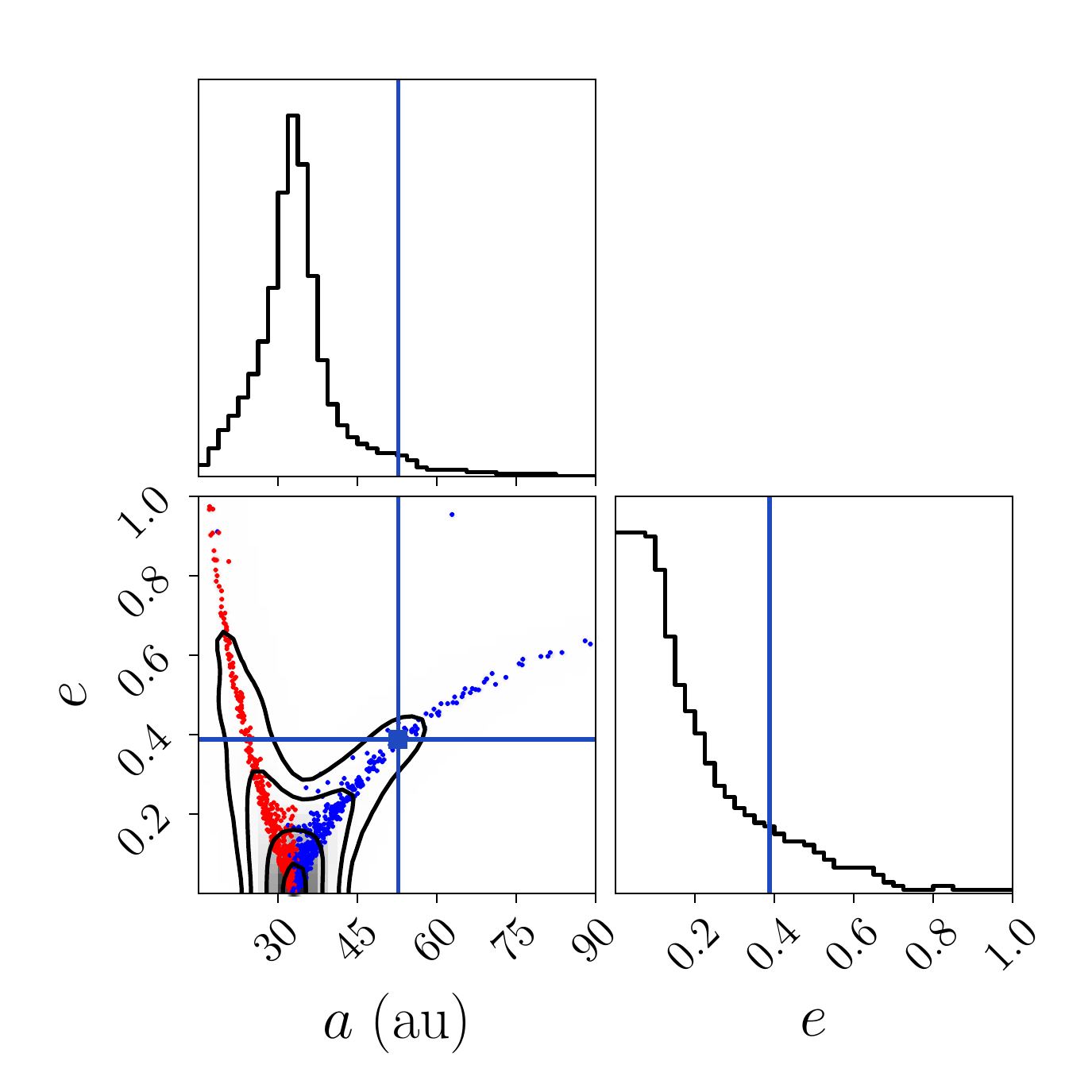}
    \caption{Zoom in of corner plot of the semi-major axis vs eccentricity solutions for system T25. The blue points on the bottom left plot represent all systems for which the mean anomaly is $M<90^\circ$ or $M>270^\circ$ (i.e. the star is closer to periastron) and the red points for $90^\circ<M<270^\circ$ (the star is closer to apastron).}
    \label{fig:T25-zoom}
\end{figure}

Some of the more subtle interesting features of this system become apparent when we examine the corner plot. The PDF of $a_\mathrm{in}$ and $e_\mathrm{in}$ is shown in Fig. \ref{fig:T25-zoom} and has an unusual structure.  There are many possible solutions for $a_\mathrm{in} \sim 20-50$au and low eccentricity, and then the possible solutions diverge into two distinct branches when $e_\mathrm{in}\gtrsim0.2$ - with fits found at low-$a$ and high-$e$, or high-$a$ and intermediate-$e$.  The possible fits have been coloured red when the system is close to apastron ($M \sim 180^{\circ}$), and blue if the system is close to periastron ($M \sim 0/360^{\circ}$).  Which `branch' is followed clearly depends on where in its orbit the system is placed.

This shows that despite the true value of the semi-major axis falls in the tail of the PDF of possible semi-major axes, it is still in a well-populated region of $a$-$e$ parameter space.  Again this shows the value of examining the corner plots rather than just relying on parameters reduced to a single dimension.

\section{Timing}
\label{subsec:timing}

Our code uses a brute-force Monte Carlo method to randomly generate fake binary or triple systems, with parameters drawn from uniform distributions (for inclination this is uniform in sin $i$). This method samples the total available parameter space as comprehensively as possible, but due to the vastness of this parameter space, we require a huge number of iterations. The code written is in {\tt fortran90} and OMP parallelised to run on multiple cores. 

The average CPU time per iteration over multiple simulations is $\sim$34 $\mathrm{ns}$, and is very similar when testing on both binary and triple systems (a typical triple system is usually rejected after only modelling the inner binary making the time per iteration very similar). 

The number of iterations required to find an appropriate number of matches varies significantly from system to system. For example, the simulation for system B38 ran for $11.2\;\mathrm{min}$ and found one match for every 42 000 fake systems tested (a match being found every $13\;\mathrm{ms}$), but system B8 ran for $35.7\;\mathrm{s}$ and found one match every $4.47\times10^6$ iterations (a match was found every $0.12\;\mathrm{ms}$). 

The majority of binary simulations have a wall-time of $\sim$1-12 min, and run for $\sim 10-160$ CPU min. The simulation that produced the results in Fig. \ref{fig:bin3-best} took 8 min 53 s to run, sampling a total of $1.9\times10^{12}$ fake systems. From these, 51 293 matches were found with separations and position angles within the errors. This corresponds to a rejection rate of over 99.99999 per cent. 

Due to the 5 additional orbital parameters that must be found to fit a triple system, the time taken to produce a sufficient number of matches for each triple simulation was significantly longer on average than for binaries. It also varied significantly from system to system, from a minimum of $\sim$2.22 mins wall-time, to less than 300 matches being found in 24 hrs of wall-time.

\section{Multi-Epoch Observations}

\begin{figure}
    \centering
    \includegraphics[width=\columnwidth]{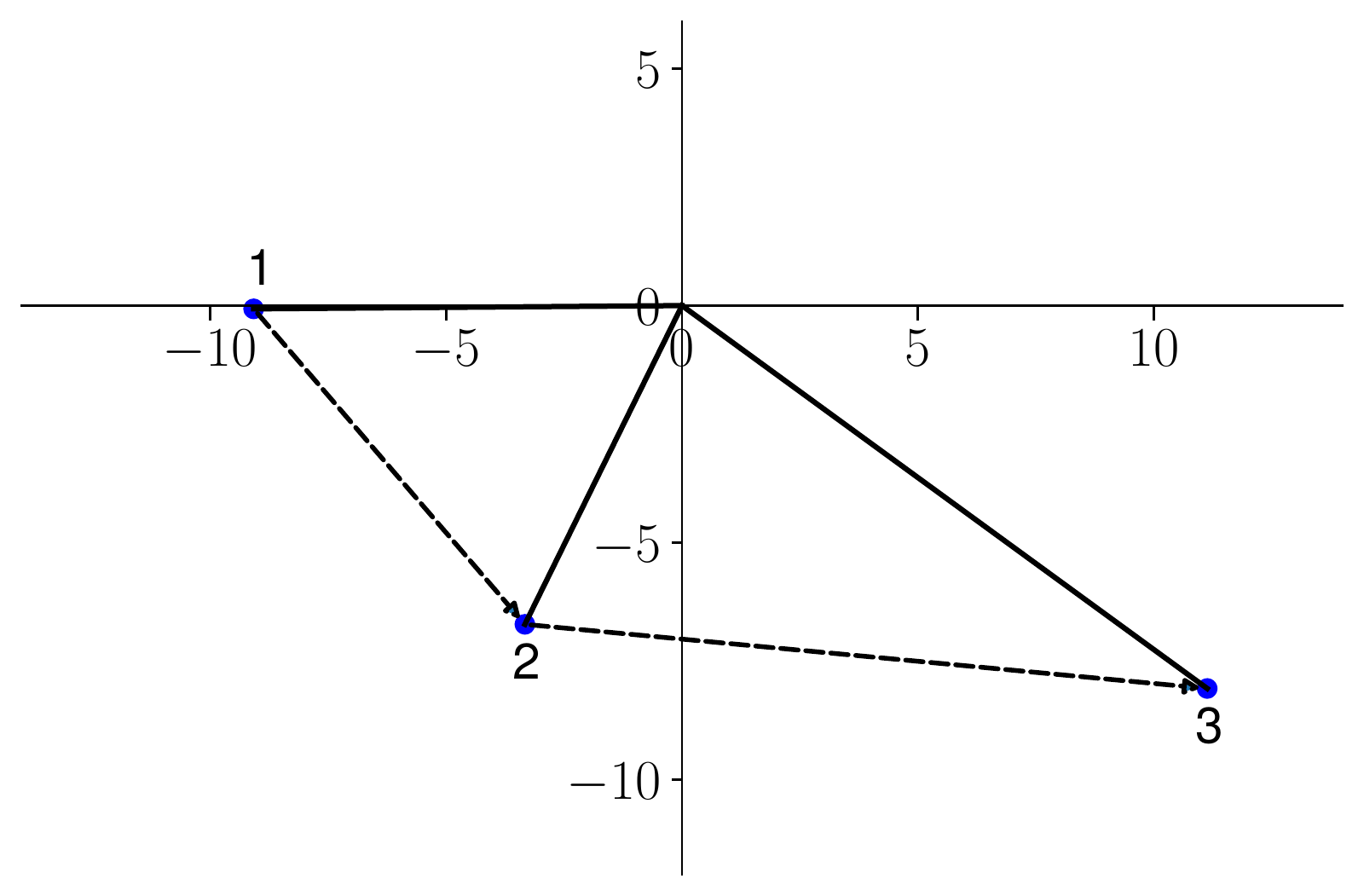}
    \caption{On sky projection of a binary system with three epochs of observation. The position of the star at the first, second and third epoch is marked by the numbers 1, 2 and 3 respectively. The primary star is centred on (0,0) for all observations and the axes given in au.}
    \label{fig:projection-me5}
\end{figure}

We have concentrated above on estimating the orbital parameters from a bare minimum of data in just two epochs of observation. However, extra information from a third epoch can sometimes (unsurprisingly) significantly improve our estimates. With more than two epochs of data we go through the procedure outlined above to fit the first two epochs, and then repeat to fit any further epochs.

\begin{figure*}
    \centering
    \includegraphics[width=.95\textwidth]{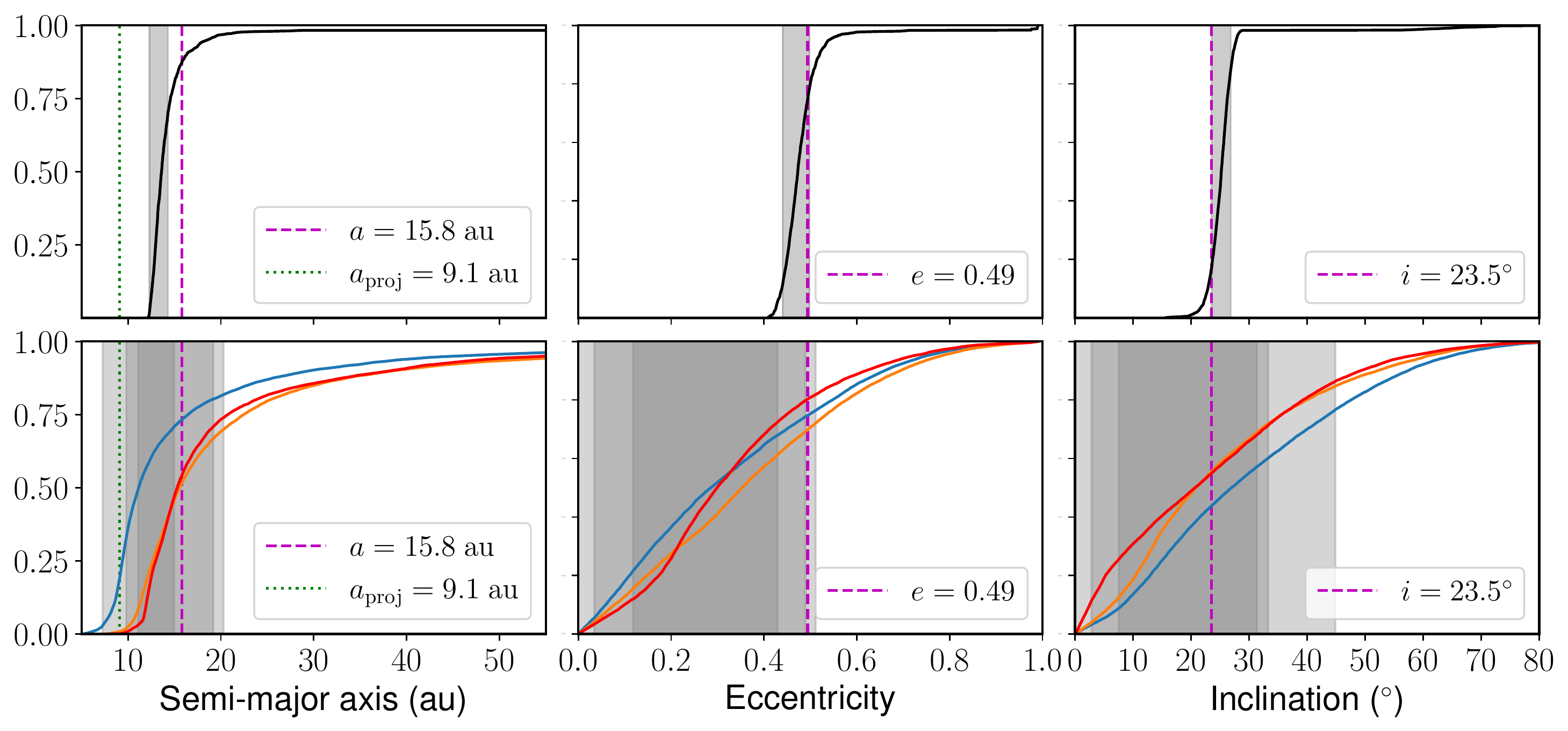}
    \caption{Cumulative distribution for a binary system with 3 epochs of observation. The top three plots show the PDFs using all three epochs together. The bottom plots show the PDFs from each possible pair of observations (blue: epochs 1 and 2, orange: epochs 2 and 3, and red: epochs 1 and 3).  The grey shaded regions represent the 68 per cent confidence interval for each CDF.}
    \label{fig:me-CDFs}
\end{figure*}

We tested the code on 20 additional fake systems with three epochs of observations each. This showed that an additional data point can sometimes be very constraining (not always, sometimes a third epoch makes very little difference). The observation of one such system is shown in Fig. \ref{fig:projection-me5}. The secondary has moved a significant distance between each observation suggesting we are seeing a reasonable fraction of its orbit (and that its period is not too many times greater than the time between epochs).

This system had a maximum projected separation of 13.8 au, from which the lower limit on the semi-major axis was calculated as $a_\mathrm{min}=6.9$ au. The three epochs of observation were separated by 7.31 and 11.45 yrs (so it was observed over an 18.76 yr timescale), giving the companion star an observed on sky velocity of $v_\mathrm{obs}=$1.20 au$\;\mathrm{yr^{-1}}$ ($5.72\;\mathrm{km\;s^{-1}}$) between the first and second epochs and $v_\mathrm{obs}=$1.27 au$\;\mathrm{yr^{-1}}$ ($6.03\;\mathrm{km\;s^{-1}}$) between the second and third epochs. The upper limit on the semi-major axis for this system was therefore $a_\mathrm{max}=3\;764$ au.

In Fig. \ref{fig:me-CDFs} we show the PDFs (as CDFs) for the semi-major axis, eccentricity, and inclinations of the system using all three epochs (top row), and using each pair of epochs (bottom row). The true values are given by the red dashed lines, and the 68 per cent confidence limits by the greyed regions. The projected separation is shown by the green dotted line for the semi-major axis.

\begin{figure}
    \centering
    \includegraphics[width=\columnwidth]{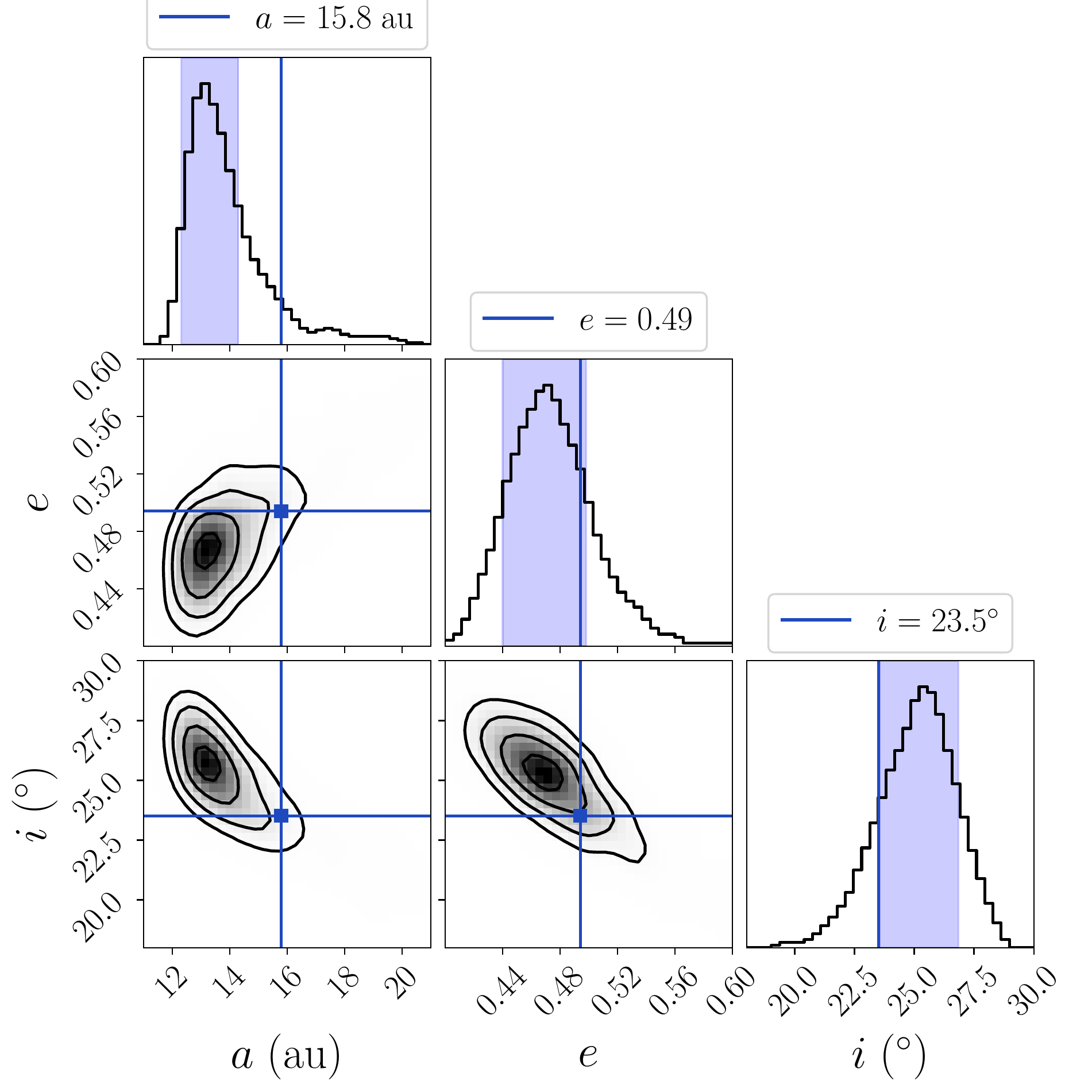}
    \caption{Corner plot of the semi-major axis, eccentricity and inclination for the system shown in Fig. \ref{fig:projection-me5}. The true value of each parameter is indicated by the solid vertical blue line.}
    \label{fig:corner-me5}
\end{figure}

The most striking feature of Fig. \ref{fig:me-CDFs} is how much a third epoch is able to constrain all three orbital parameters in this case. Fig. \ref{fig:corner-me5} shows the corner plot of semi-major axis, eccentricity and inclination, with histograms featured in the top plot of each column and parameter covariances shown in the other panels. This highlights how tightly each parameter is constrained using the three epoch method when one sees how small the ranges of $a$, $e$, and $i$ are.

The 68 per cent confidence limits on the semi-major axis have fallen from about $8 - 20\;\mathrm{au}$ to $12.3 - 14.3\;\mathrm{au}$. The true value of the semi-major axis for this system is $15.8\;\mathrm{au}$, falling outside the 68 per cent confidence interval but within the 95 per cent confidence interval of $12.1 - 18.6\;\mathrm{au}$.

Similarly, the inclination true inclination of $23.5^{\circ}$ falls at the lower end of the 68 per cent confidence interval ($23.5 - 26.8^{\circ}$) and comfortably within the 95 per cent limits of $21.0 - 28.3^{\circ}$. The true eccentricity value of 0.49 falls within both the 68 per cent (0.44 - 0.50) and 95 per cent (0.41 - 0.54) confidence intervals. 

The reason an extra epoch is so much more constraining for this system is that we have three epochs spanning $\sim 19$ yrs of a $\sim 50$ yr total period.  Hence the third epoch requires a large on-sky motion in a very particular direction from any fits to the first two epochs which `pins down' the orbit extremely well.  When we test on systems where three epochs only cover a small fraction of an orbit and have large observational errors we find that the third epoch can sometimes add very little to the constraints from just two.

\section{Comparisons}
\label{sec:comparisons}

It is worth comparing {\tt FOBOS} to some other orbit-fitting codes.  Note that {\tt FOBOS} is deliberately designed to be used in situations where we have minimal astrometric data only.  Other codes are often designed to use many more epochs with extra (e.g. velocity) information gained from a sustained and detailed observing program. If such additional data exists we suggest using these codes rather than {\tt FOBOS}.

We used {\tt FOBOS} to constrain the orbits of several observed binary systems and compared our results to various Bayesian Markov Chain Monte Carlo fitting methods. In this section, we present our results for the binary systems 2MASS J01033563-5515561 (\cite{OFTI}) and HD 206893 B (\cite{Ward-Duong2021}), using two epochs of astrometric observations for the 2MASS binary and four epochs of astrometric observations for HD 206893 B.

The true observational errors on the separations, position angles, and distances were used to determine whether a particular set of orbital parameters is a match to the observations. The impact of the size of the observational errors is discussed later in this section.

\subsubsection{2M 0103-55 (AB) b}

2MASS J01033563-5515561 (AB) b (hereafter 2M 0103-55 (AB) b) is a 12-14 Jupiter mass companion orbiting the low mass binary system 2M 0103-55 (AB). The astrometric data for this system was acquired by \cite{Delorme+2013} and analysed using the Orbits for the Impatient (OFTI) method \citep{OFTI}. \cite{OFTI} used two epoch of relative astrometry taken $\sim$10 years apart (see their Table 10) to generate the orbital parameter posteriors for 2M 0103-55 (AB) b. The same two astrometric data points were used as the input for {\tt FOBOS}. 

The separations quoted in this table are measured relative to the barycentre of the system 2M 0103-55 (AB). The errors on the position angles (PA) corresponds to the relative error on the observations between the two epochs, and both PA measurements have an additional error of $\pm0.4$, dominated by systematic uncertainties. A distance of $d=47.2\pm3.1\;\rm{pc}$ (obtained using the parallax quoted in \cite{OFTI} Table 2) was used to convert the separations from milliarcseconds to au. The masses of the host binary system (treated as a single object) and the low mass companion were taken to be $M_{AB}=0.36\pm0.04\;\rm{M_\odot}$ and $M_b=0.012\pm0.001\;\rm{M_\odot}$ respectively. 

\begin{figure}
    \centering
    \includegraphics[width=\columnwidth]{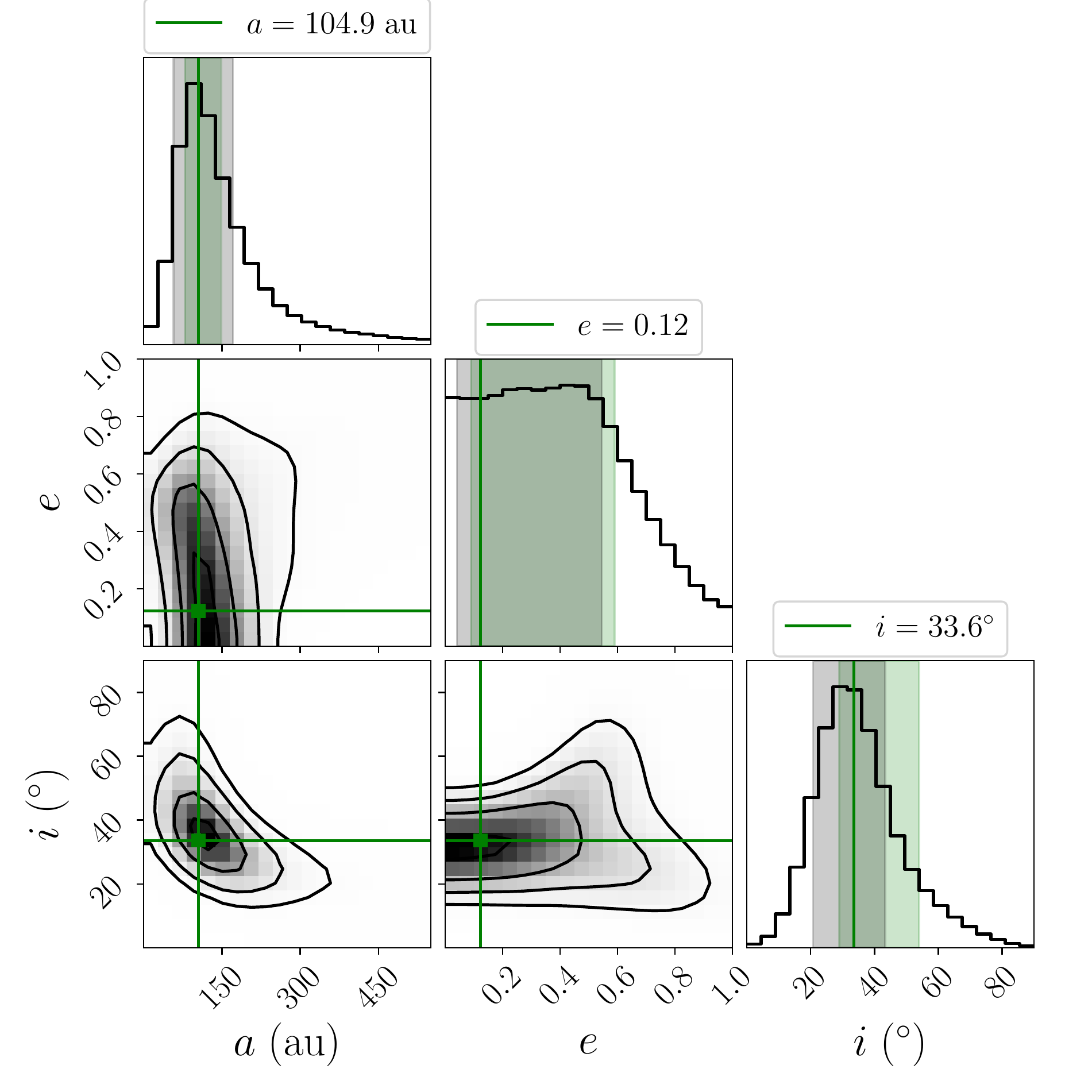}
    \caption{Corner plot for the orbit of 2M 0103-55 (AB) b with respect to 2M 0103-55 (AB). The top panels of each column show the {\tt FOBOS} probability distribution functions for semi-major axis (left), eccentricity (middle), and inclination (right). The green solid lines on these panels show the OFTI median values and the green shades regions show their 68 per cent confidence intervals (\citet{OFTI}, Table 20). The grey shaded regions are the {\tt FOBOS} 68 per cent confidence intervals.}
    \label{fig:corner-2M010355}
\end{figure}

Using the Orbits for the Impatient algorithm, \cite{OFTI} find median values for the semi-major axis, eccentricity, and inclination to be $a=104.92\;\rm{au}$, $e=0.1233$, and $i=123.6^\circ$, and 68 per cent confidence intervals of $79-149\;\rm{au}$, $0.09-0.59$, and $119-144^\circ$, measured relative to the system being edge-on at $90^\circ$. As mentioned earlier, {\tt FOBOS} defines edge-on as $0^\circ$, so this corresponds to an inclination range of $29-54^\circ$ using the {\tt FOBOS} frame of reference. These results are shown in Fig. \ref{fig:corner-2M010355} by the green vertical lines and green shaded regions respectively. 

{\tt FOBOS} calculates the 68 per cent confidence intervals for the semi-major axis, eccentricity, inclination as $59.1 - 173.8\;\mathrm{au}$, $0.01 - 0.52$, and $19.9 - 44.7^\circ$ respectively; these ranges are indicated on Fig. \ref{fig:corner-2M010355} by the grey shaded regions. The median values for all three orbital parameters fall within the {\tt FOBOS} 68 per cent confidence intervals and we see a significant overlap between all of the {\tt FOBOS} and OFTI 68 per cent confidence intervals. 

The widths of the confidence intervals for eccentricity ($\sim0.5$) and inclination ($\sim25^\circ$) calculated using {\tt FOBOS} match those quoted by \cite{OFTI}, but the {\tt FOBOS} semi-major axis range is $\sim1.6$ times larger than the OFTI range. For the inclination, the 68 per cent C. I. is a comparable width to that calculated by OFTI, but shifted to slightly lower values. 

The {\tt FOBOS} simulation of 2M 0103-55 produces over 50,000 solutions within the 1$\sigma$ observational errors calculated by \cite{Delorme+2013} in a wall-time of $\sim$30 seconds. 

\subsubsection{HD 206893 B}

Further tests were carried out on HD 206893 B - a 12-40 Jupiter mass companion orbiting in the debris disk of its FV5 type host star. A detailed analysis of the physical and orbital properties of HD 206893 B was presented in \cite{Ward-Duong2021}, using a total of nine astrometric observations from previous VLT/SPHERE, VLT/NaCo studies of the system \citep{Milli+2017,Delorme+2017,Grandjean+2019} and new Gemini Planet Imager \citep{GPI} observations. These data points are given in Table 9 of \cite{Ward-Duong2021}. 

HD 206893 B has a Gaia distance of $d=40.77\pm0.059\;\rm{pc}$ and the host star and companion star have masses of $M_A=1.31\pm0.01\;\rm{M_\odot}$ and $M_B=0.11\pm0.01\;\rm{M_\odot}$ respectively. 

Using a Bayesian MCMC method, \cite{Ward-Duong2021} find the semi-major axis of the system to be $10.4^{+1.8}_{-1.7}\;\mathrm{au}$ and an eccentricity of $0.23^{+0.13}_{-0.16}$. 

\begin{figure}
    \centering
    \includegraphics[width=\columnwidth]{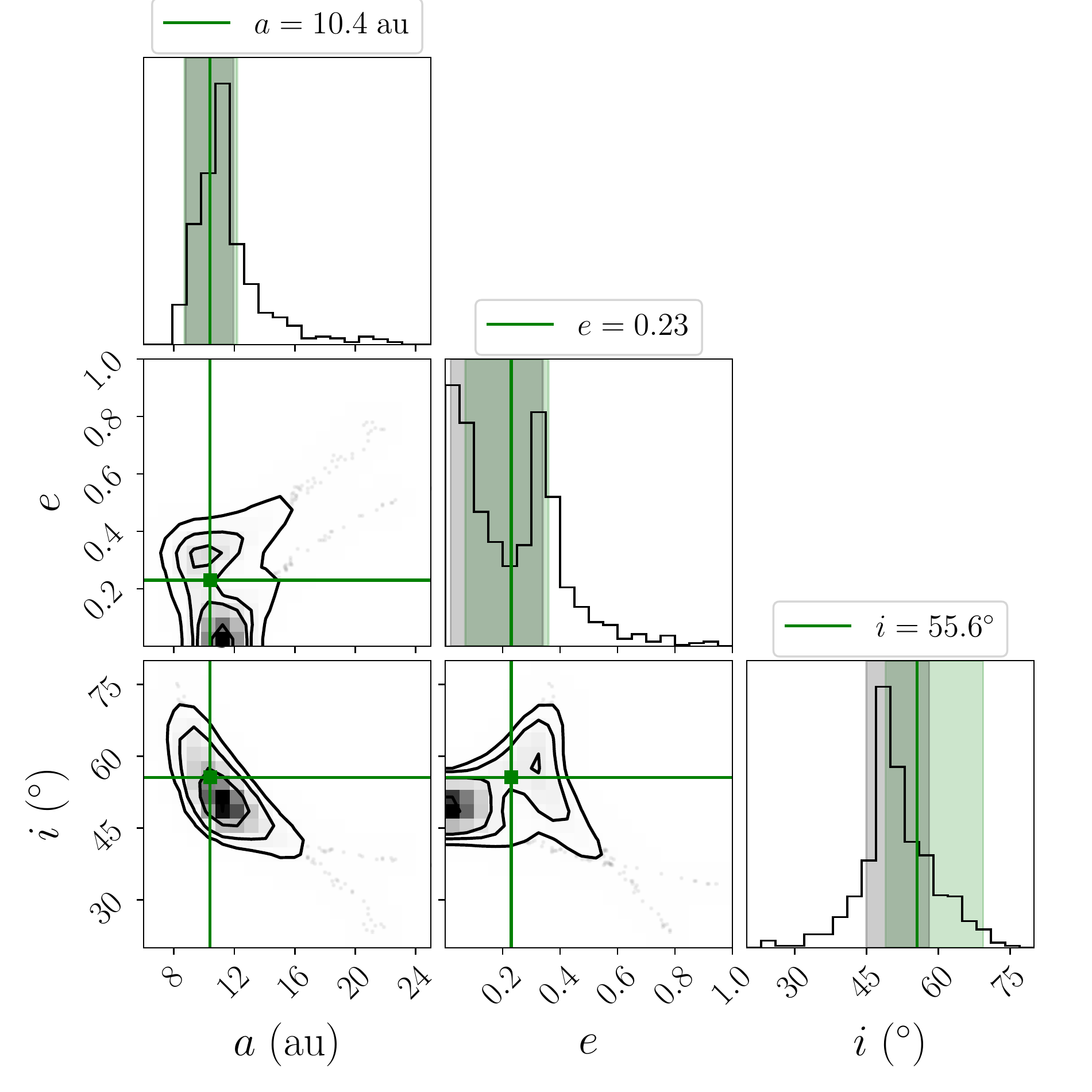}
    \caption{Corner plot showing the probability density functions for semi-major axis (top left), eccentricity (top middle), and inclination (right) for the low mass companion HD 206893 B. The other panels in the figure show the covariances of each of these parameters. The solid green lines show the most probable values for each orbital parameter obtained by \citet{Ward-Duong2021} and the green shaded regions represent their $1\sigma$ error ranges. The grey shaded regions are the {\tt FOBOS} 68 per cent confidence intervals.}
    \label{fig:corner_HD206893B}
\end{figure}

They also find a inclination of $145.6\degr^{+13.8\degr}_{-6.6\degr}$, corresponding to $55.6\degr^{+13.8\degr}_{-6.6\degr}$ using our definition. Their most probable values and 1$\sigma$ ranges are shown on Fig. \ref{fig:corner_HD206893B} by the green vertical lines and shaded regions, with the inclination values being shifted down by 90$^\circ$ to match our definition of inclination. The corner plot showing their posterior distributions and covariances is shown in their Table 10. 

\begin{figure}
    \centering
    \includegraphics[width=\columnwidth]{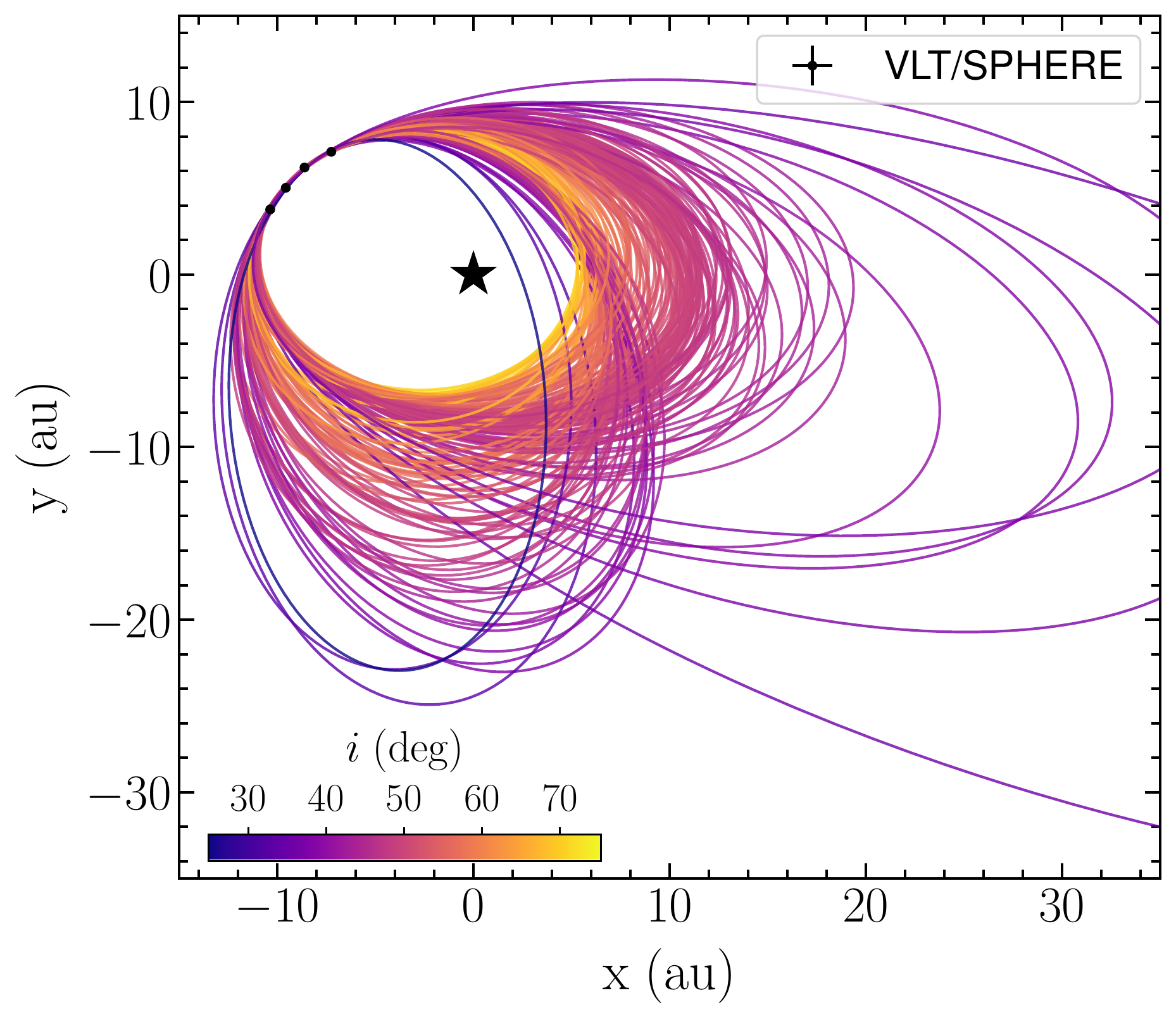}
    \caption{Subsample of 200 randomly selected orbital fits for HD 206893 B using the four epochs of VLT/SPHERE astrometric observations (black circles). The observational errors on the measurements are plotted as error bars on the black points, but are too small to be noticeable on this scale. The lighter/yellower orbits correspond to systems with inclinations closer to face-on (90$^\circ$) and the darker/bluer orbits closer to edge-on (0$^\circ$). The primary star is located at (0,0).}
    \label{fig:orbits_HD206893B}
\end{figure}

We tested the {\tt FOBOS} Multi-Epoch code on the four SPHERE/IRDIS observations. Using a 6 core/12 thread processor, $\sim$1,000 matches to the observations for this system are found in a wall-time of $\sim60$ minutes. These results were used to generate the probability distribution functions (top panels) and covariances (other panels) shown in Fig. \ref{fig:corner_HD206893B}. 

We calculate the 68 per cent confidence intervals for semi-major axis, eccentricity, and inclination as $8.8-11.9\;\mathrm{au}$, $0.02-0.34$, and $45.0-58.1^\circ$. These ranges are represented by the grey shaded regions on the top panels of Fig. \ref{fig:corner_HD206893B}. The top left panel of the plot shows that our confidence interval for $a$ overlaps with the range from \cite{Ward-Duong2021} almost exactly. The {\tt FOBOS} confidence interval extends to slightly lower values than the Ward-Duong CI and the PDF shows the same minimum at $\sim$0.2 followed by a peak at $\sim$0.3, before tailing off almost completely for values $\gtrsim$0.5. The {\tt FOBOS} 68 per cent CI for inclination is 1.5 times smaller than the \cite{Ward-Duong2021} value and shifted to a slightly smaller inclination range, with the median value falling in the region where the two ranges overlap. 

A sample of 200 orbits which fit the four VLT/SPHERE observations are shown in Fig. \ref{fig:orbits_HD206893B}. The colour of the orbit represents whether the inclination of HD 206893 B is closer to edge-on (0$^\circ$, bluer orbits) or face-on (90$^\circ$). 

\subsubsection{Observational errors}
\label{subsubsec:obserrors}

Comparing the results for 2M 0103-55 (AB) b and HD 206893 B, we see that HD 206893 B is much more highly constrained by {\tt FOBOS}. This is almost completely due to the additional epochs of data available for HD 206893 B. However, we also found a difference in results depending on whether the four GPI observations or the four SPHERE observations were used. Fitting the GPI observations resulted in a semi-major axis confidence interval that was $\sim$2.5 times larger than the equivalent results using the SPHERE observations, and a $\sim$1.7 times increase in the inclination range. 

There are two reasons why the VLT/SPHERE observations are much better at constraining the orbital parameters than the GPI observations. Firstly, two of the GPI observations were obtained within one month of each other and their 1$\sigma$ error ranges overlap for both separation and position angle. Secondly, the fourth data point has 1$\sigma$ errors that are $\sim$2 times larger than the errors for all other data points. This emphasises the importance of obtaining data points with small observational errors over a long enough timescale that we see the companion exhibit significant on sky motion.

\section{Conclusions}

The Few Observation Binary Orbit Solver ({\tt FOBOS}) is a (usually) very fast way of finding statistically reliable confidence limits on the orbital parameters of binary and triple systems from as little as two epochs of purely astrometric data. This allows orbital parameters to be estimated from limited astrometric data (such as from follow-up observations on systems) from which it might not have previously been considered possible to extract estimates of the orbital parameters.  

{\tt FOBOS} uses a brute force Monte Carlo approach with flat priors to search parameter space for solutions that produce fits to within the observational errors of the astrometric observations. It can find significant numbers of possible matches usually in a few CPU minutes for binary systems, or a few CPU hours for triple systems.

How constraining fits are is a matter of `luck' in that some pairs of observations can be very constraining, while others may contain little information. Unsurprisingly, smaller observational errors usually allow {\tt FOBOS} to be more constraining. The 68 and 95 per cent confidence limits are statistically reliable (and so tight constraints can be statistically trusted).

{\tt FOBOS} has been tested on a large sample of fake binary and triple samples to prove its reliability for systems with known parameters. We have also tested {\tt FOBOS} on two observed binary systems, showing that the results generally agree with fits from other well-established orbital fitting codes.

{\tt FOBOS} has applications in finding orbital solutions for binary and triple systems studied as part of multiplicity surveys, and can also be applied to directly imaged exoplanets. 

\section*{Acknowledgements}

We thank the anonymous referee for helpful feedback that improved the quality of this work. RJH acknowledges support from the UK Science and Technology Facilities Council in the form of a PhD studentship. RJH also thanks Alex Brown, Luke Holden and Richard Parker for helpful discussions and feedback. For the purpose of open access, the author has applied a CC BY public copyright licence to any Author Accepted Manuscript version arising.

\section*{Data Availability}

The FOBOS orbital fitting code is hosted on GitHub\footnote{\url{https://www.github.com/rebeccahoughton/FOBOS}}. All data underlying this article will be shared upon reasonable request.



\bibliographystyle{mnras}
\bibliography{paper} 

\begin{thebibliography}{}
\makeatletter
\relax
\def\mn@urlcharsother{\let\do\@makeother \do\$\do\&\do\#\do\^\do\_\do\%\do\~}
\def\mn@doi{\begingroup\mn@urlcharsother \@ifnextchar [ {\mn@doi@}
  {\mn@doi@[]}}
\def\mn@doi@[#1]#2{\def\@tempa{#1}\ifx\@tempa\@empty \href
  {http://dx.doi.org/#2} {doi:#2}\else \href {http://dx.doi.org/#2} {#1}\fi
  \endgroup}
\def\mn@eprint#1#2{\mn@eprint@#1:#2::\@nil}
\def\mn@eprint@arXiv#1{\href {http://arxiv.org/abs/#1} {{\tt arXiv:#1}}}
\def\mn@eprint@dblp#1{\href {http://dblp.uni-trier.de/rec/bibtex/#1.xml}
  {dblp:#1}}
\def\mn@eprint@#1:#2:#3:#4\@nil{\def\@tempa {#1}\def\@tempb {#2}\def\@tempc
  {#3}\ifx \@tempc \@empty \let \@tempc \@tempb \let \@tempb \@tempa \fi \ifx
  \@tempb \@empty \def\@tempb {arXiv}\fi \@ifundefined
  {mn@eprint@\@tempb}{\@tempb:\@tempc}{\expandafter \expandafter \csname
  mn@eprint@\@tempb\endcsname \expandafter{\@tempc}}}

\bibitem[\protect\citeauthoryear{{Blunt} et~al.,}{{Blunt} et~al.}{2017}]{OFTI}
{Blunt} S.,  et~al., 2017, \mn@doi [\aj] {10.3847/1538-3881/aa6930}, \href
  {https://ui.adsabs.harvard.edu/abs/2017AJ....153..229B} {153, 229}

\bibitem[\protect\citeauthoryear{{Blunt} et~al.,}{{Blunt}
  et~al.}{2020}]{Orbitize}
{Blunt} S.,  et~al., 2020, \mn@doi [\aj] {10.3847/1538-3881/ab6663}, \href
  {https://ui.adsabs.harvard.edu/abs/2020AJ....159...89B} {159, 89}

\bibitem[\protect\citeauthoryear{{Delorme} et~al.,}{{Delorme}
  et~al.}{2013}]{Delorme+2013}
{Delorme} P.,  et~al., 2013, \mn@doi [\aap] {10.1051/0004-6361/201321169},
  \href {https://ui.adsabs.harvard.edu/abs/2013A&A...553L...5D} {553, L5}

\bibitem[\protect\citeauthoryear{{Delorme} et~al.,}{{Delorme}
  et~al.}{2017}]{Delorme+2017}
{Delorme} P.,  et~al., 2017, \mn@doi [\aap] {10.1051/0004-6361/201731145},
  \href {https://ui.adsabs.harvard.edu/abs/2017A&A...608A..79D} {608, A79}

\bibitem[\protect\citeauthoryear{{Duch{\^e}ne} \& {Kraus}}{{Duch{\^e}ne} \&
  {Kraus}}{2013}]{DucheneKraus2013}
{Duch{\^e}ne} G.,  {Kraus} A.,  2013, \mn@doi [\araa]
  {10.1146/annurev-astro-081710-102602}, \href
  {https://ui.adsabs.harvard.edu/abs/2013ARA&A..51..269D} {51, 269}

\bibitem[\protect\citeauthoryear{{Eggleton} \& {Kiseleva}}{{Eggleton} \&
  {Kiseleva}}{1995}]{EggletonKiseleva1995}
{Eggleton} P.,  {Kiseleva} L.,  1995, \mn@doi [\apj] {10.1086/176611}, \href
  {https://ui.adsabs.harvard.edu/abs/1995ApJ...455..640E} {455, 640}

\bibitem[\protect\citeauthoryear{{Fulton}, {Petigura}, {Blunt}  \&
  {Sinukoff}}{{Fulton} et~al.}{2018}]{RadVel}
{Fulton} B.~J.,  {Petigura} E.~A.,  {Blunt} S.,   {Sinukoff} E.,  2018, \mn@doi
  [\pasp] {10.1088/1538-3873/aaaaa8}, \href
  {https://ui.adsabs.harvard.edu/abs/2018PASP..130d4504F} {130, 044504}

\bibitem[\protect\citeauthoryear{{Goodwin}}{{Goodwin}}{2010}]{Goodwin2010}
{Goodwin} S.~P.,  2010, \mn@doi [Philosophical Transactions of the Royal
  Society of London Series A] {10.1098/rsta.2009.0254}, \href
  {https://ui.adsabs.harvard.edu/abs/2010RSPTA.368..851G} {368, 851}

\bibitem[\protect\citeauthoryear{{Grandjean} et~al.,}{{Grandjean}
  et~al.}{2019}]{Grandjean+2019}
{Grandjean} A.,  et~al., 2019, \mn@doi [\aap] {10.1051/0004-6361/201935044},
  \href {https://ui.adsabs.harvard.edu/abs/2019A&A...627L...9G} {627, L9}

\bibitem[\protect\citeauthoryear{{Harrington}}{{Harrington}}{1972}]{Harrington1972}
{Harrington} R.~S.,  1972, \mn@doi [Celestial Mechanics] {10.1007/BF01231475},
  \href {https://ui.adsabs.harvard.edu/abs/1972CeMec...6..322H} {6, 322}

\bibitem[\protect\citeauthoryear{{Kreidberg}}{{Kreidberg}}{2015}]{BATMAN}
{Kreidberg} L.,  2015, \mn@doi [\pasp] {10.1086/683602}, \href
  {https://ui.adsabs.harvard.edu/abs/2015PASP..127.1161K} {127, 1161}

\bibitem[\protect\citeauthoryear{{Macintosh} et~al.,}{{Macintosh}
  et~al.}{2008}]{GPI}
{Macintosh} B.~A.,  et~al., 2008, in {Hubin} N.,  {Max} C.~E.,   {Wizinowich}
  P.~L.,  eds,  Society of Photo-Optical Instrumentation Engineers (SPIE)
  Conference Series Vol. 7015, Adaptive Optics Systems. p. 701518,
  \mn@doi{10.1117/12.788083}

\bibitem[\protect\citeauthoryear{{Mardling} \& {Aarseth}}{{Mardling} \&
  {Aarseth}}{1999}]{MardlingAarseth1999}
{Mardling} R.,  {Aarseth} S.,  1999, in {Steves} B.~A.,  {Roy} A.~E.,  eds,
  NATO Advanced Study Institute (ASI) Series C Vol. 522, The Dynamics of Small
  Bodies in the Solar System, A Major Key to Solar System Studies. p.~385

\bibitem[\protect\citeauthoryear{{Mede} \& {Brandt}}{{Mede} \&
  {Brandt}}{2017}]{ExoSOFT}
{Mede} K.,  {Brandt} T.~D.,  2017, \mn@doi [\aj] {10.3847/1538-3881/aa5e4a},
  \href {https://ui.adsabs.harvard.edu/abs/2017AJ....153..135M} {153, 135}

\bibitem[\protect\citeauthoryear{{Milli} et~al.,}{{Milli}
  et~al.}{2017}]{Milli+2017}
{Milli} J.,  et~al., 2017, \mn@doi [\aap] {10.1051/0004-6361/201629908}, \href
  {https://ui.adsabs.harvard.edu/abs/2017A&A...597L...2M} {597, L2}

\bibitem[\protect\citeauthoryear{{Reipurth} \& {Mikkola}}{{Reipurth} \&
  {Mikkola}}{2012}]{ReipurthMikkola2012}
{Reipurth} B.,  {Mikkola} S.,  2012, \mn@doi [\nat] {10.1038/nature11662},
  \href {https://ui.adsabs.harvard.edu/abs/2012Natur.492..221R} {492, 221}

\bibitem[\protect\citeauthoryear{{Reipurth}, {Clarke}, {Boss}, {Goodwin},
  {Rodr{\'\i}guez}, {Stassun}, {Tokovinin}  \& {Zinnecker}}{{Reipurth}
  et~al.}{2014}]{ReipurthPPVI}
{Reipurth} B.,  {Clarke} C.~J.,  {Boss} A.~P.,  {Goodwin} S.~P.,
  {Rodr{\'\i}guez} L.~F.,  {Stassun} K.~G.,  {Tokovinin} A.,   {Zinnecker} H.,
  2014, in {Beuther} H.,  {Klessen} R.~S.,  {Dullemond} C.~P.,   {Henning} T.,
  eds, Protostars and Planets VI. p.~267 (\mn@eprint {arXiv} {1403.1907}),
  \mn@doi{10.2458/azu\_uapress\_9780816531240-ch012}

\bibitem[\protect\citeauthoryear{{Valtonen}, {Myll{\"a}ri}, {Orlov}  \&
  {Rubinov}}{{Valtonen} et~al.}{2008}]{valtonen2008}
{Valtonen} M.,  {Myll{\"a}ri} A.,  {Orlov} V.,   {Rubinov} A.,  2008, in
  {Vesperini} E.,  {Giersz} M.,   {Sills} A.,  eds,  IAU Symposium Vol. 246,
  Dynamical Evolution of Dense Stellar Systems. pp 209--217,
  \mn@doi{10.1017/S1743921308015627}

\bibitem[\protect\citeauthoryear{{Ward-Duong} et~al.,}{{Ward-Duong}
  et~al.}{2021}]{Ward-Duong2021}
{Ward-Duong} K.,  et~al., 2021, \mn@doi [\aj] {10.3847/1538-3881/abc263}, \href
  {https://ui.adsabs.harvard.edu/abs/2021AJ....161....5W} {161, 5}

\bibitem[\protect\citeauthoryear{{Winn} \& {Fabrycky}}{{Winn} \&
  {Fabrycky}}{2015}]{Winn2015}
{Winn} J.~N.,  {Fabrycky} D.~C.,  2015, \mn@doi [\araa]
  {10.1146/annurev-astro-082214-122246}, \href
  {https://ui.adsabs.harvard.edu/abs/2015ARA&A..53..409W} {53, 409}

\makeatother
\end{thebibliography}



\appendix
\section{Additional covariances}
\label{sec:appendix1}

\begin{figure*}
    \centering
    \includegraphics[width=\textwidth]{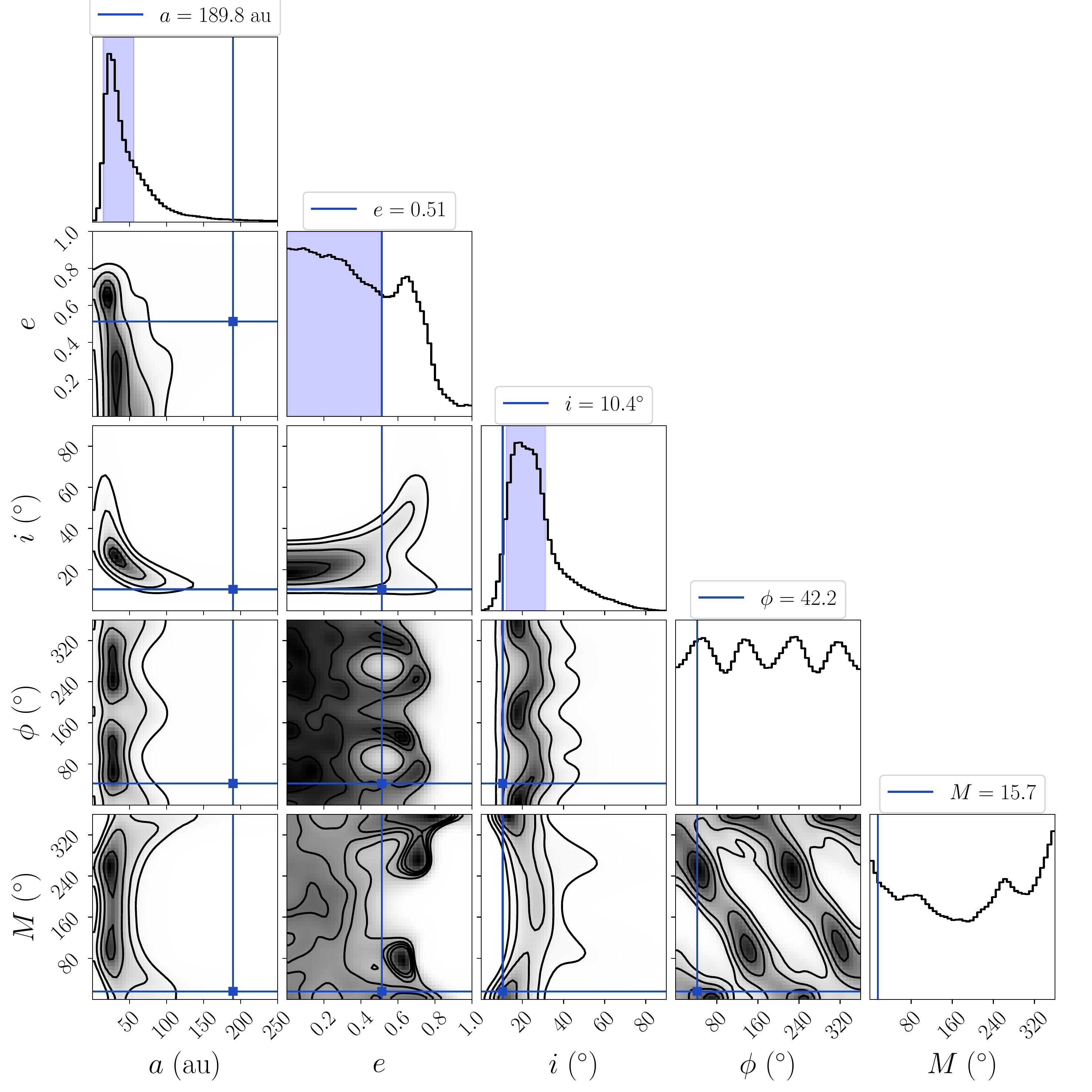}
    \caption{Corner plot showing orbital parameter covariances for test system B4. See Fig. \ref{fig:bin-corner3}.}
    \label{fig:bin-corner4}
\end{figure*}

\begin{figure*}
    \centering
    \includegraphics[width=\textwidth]{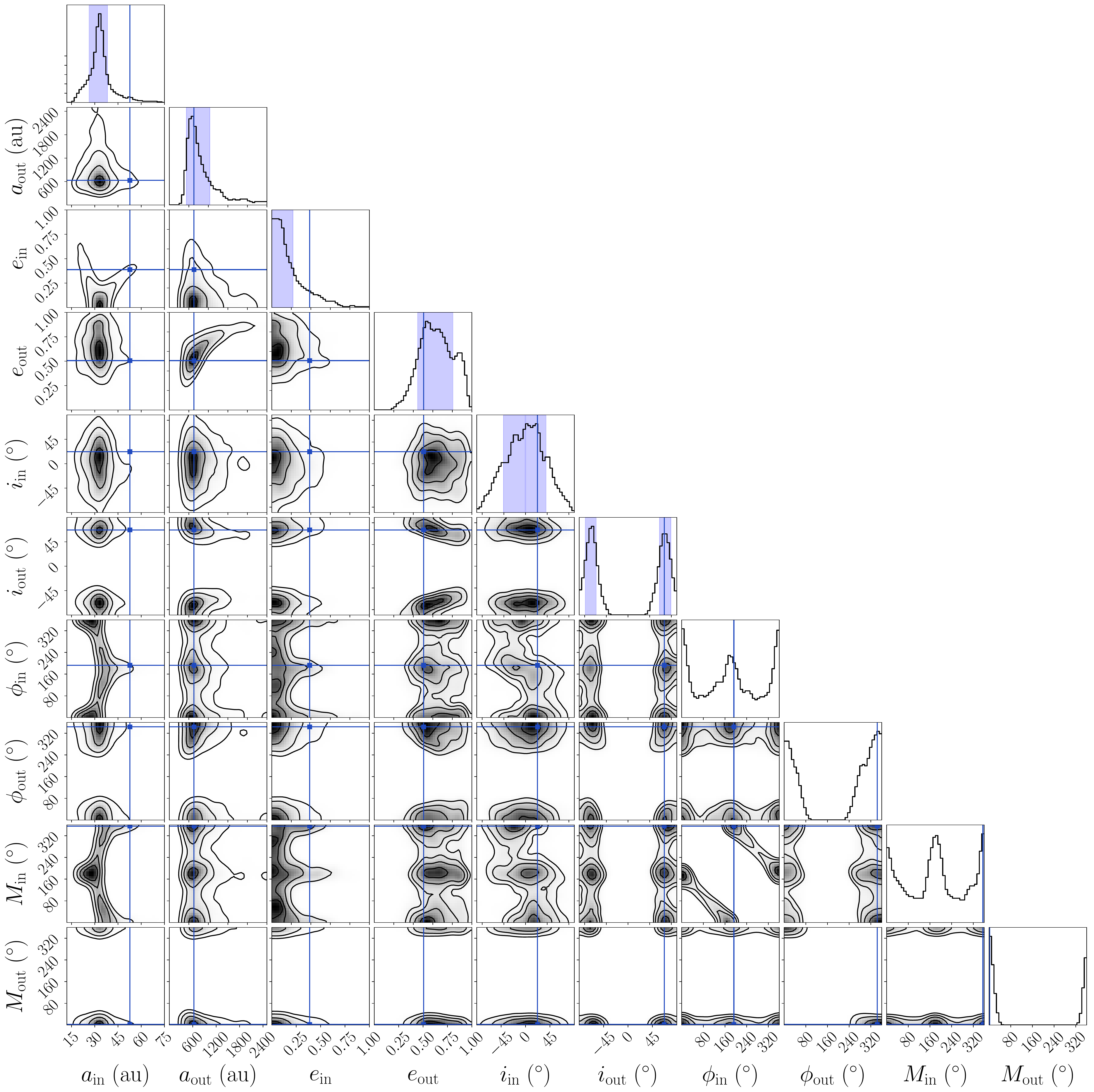}
    \caption{Corner plot for triple system 25. Sample size of 1011 matches. See Fig. \ref{fig:corner-trip19}.}
    \label{fig:corner-trip25}
\end{figure*}

Fig.  \ref{fig:bin-corner4} shows the probability distribution functions (top panels of each column) and parameter covariances for test system B4, and Fig. \ref{fig:corner-trip25} shows these properties for system T25. 

\bsp	
\label{lastpage}
\end{document}